\documentclass[longauth]{aa} 
\usepackage{graphicx}
\graphicspath{{figures/}}
\usepackage{amsmath}
\usepackage{txfonts}
\usepackage{fourier}
\usepackage{url}
\usepackage{longtable}
\usepackage{tikz}
\usepackage{chemformula}
\usepackage{natbib}
\usepackage{subfig}
\usepackage[
bookmarks=true,           
bookmarksnumbered=true,   
colorlinks=true,          
citecolor=blue,           
linkcolor=blue,           
menucolor=blue,           
urlcolor=blue,            
linkbordercolor={0 0 1},  
pdfborder={0 0 1},
frenchlinks=true]{hyperref}

\usepackage[flushleft]{threeparttable}

\DeclareSymbolFont{UPM}{U}{eur}{m}{n}
\DeclareMathSymbol{\umu}{0}{UPM}{"16}
\let\oldumu=\umu
\renewcommand\umu{\ifmmode\oldumu\else\math{\oldumu}\fi}
\newcommand\micro{\umu}
\newcommand\micron{\micro\rm m}
\newcommand\microns{\micron}
\newcommand\Spitzer{{\em Spitzer}}

\newcommand\HST{{\em HST}}

\newcommand\pyratbay{\textsc{Pyrat Bay}}
\newcommand\repack{\textsc{repack}}

\newcommand\molhyd{\ifmmode{{\rm H}\sb{2}}\else{H$\sb{2}$}\fi}
\newcommand\methane{\ifmmode{{\rm CH}\sb{4}}\else{CH$\sb{4}$}\fi}
\newcommand\water{\ifmmode{{\rm H}\sb{2}{\rm O}}\else{H$\sb{2}$O}\fi}
\newcommand\carbdiox{\ifmmode{{\rm CO}\sb{2}}\else{CO$\sb{2}$}\fi}
\newcommand\carbmono{\ifmmode{{\rm CO}}\else{CO}\fi}
\newcommand\ammonia{\ifmmode{{\rm NH}\sb{3}}\else{NH$\sb{3}$}\fi}
\newcommand\acetylene{\ifmmode{{\rm C}\sb{2}{\rm H}\sb{2}}
                        \else{C$\sb{2}$H$\sb{2}$}\fi}
\newcommand\ethylene{\ifmmode{{\rm C}\sb{2}{\rm H}\sb{4}}
                        \else{C$\sb{2}$H$\sb{4}$}\fi}
\newcommand\cyanide{\ifmmode{{\rm HCN}}\else{HCN}\fi}
\newcommand\nitrogen{\ifmmode{{\rm N}\sb{2}}\else{N$\sb{2}$}\fi}
\newcommand{\teff}{$T_{\rm eff}$}
\newcommand{\teqm}{$T_{\rm eq}$}

\newcommand{\keltb}{KELT-20\,b}
\newcommand{\mascarab}{MASCARA-2\,b}
\newcommand{\kelt}{KELT-20}
\newcommand{\cheops}{CHEOPS}
\newcommand{\tess}{TESS}

\newcommand{\vsini}{\ensuremath{v \sin i_\star}\xspace}

\begin{document}

\title{\textit{\cheops} observations of \keltb/\mascarab: An aligned orbit and signs of variability from a reflective day side.\thanks{This study uses \cheops\ data observed as part of the Guaranteed Time Observation (GTO) programmes CH\_PR100016, CH\_PR110047, and CH\_PR100020. The CHEOPS detrended photometry discussed in this article is available in electronic form at CDS. (link will be added here) }}

\author{
V. Singh\inst{1} $^{\href{https://orcid.org/0000-0002-7485-6309}{\includegraphics[scale=0.5]{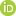}}}$, 
G. Scandariato\inst{1} $^{\href{https://orcid.org/0000-0003-2029-0626}{\includegraphics[scale=0.5]{Figures/orcid.jpg}}}$, 
A. M. S. Smith\inst{2} $^{\href{https://orcid.org/0000-0002-2386-4341}{\includegraphics[scale=0.5]{Figures/orcid.jpg}}}$, 
P. E. Cubillos\inst{3,4}, 
M. Lendl\inst{5} $^{\href{https://orcid.org/0000-0001-9699-1459}{\includegraphics[scale=0.5]{Figures/orcid.jpg}}}$, 
N. Billot\inst{5} $^{\href{https://orcid.org/0000-0003-3429-3836}{\includegraphics[scale=0.5]{Figures/orcid.jpg}}}$, 
A. Fortier\inst{6,7} $^{\href{https://orcid.org/0000-0001-8450-3374}{\includegraphics[scale=0.5]{Figures/orcid.jpg}}}$, 
D. Queloz\inst{8,9} $^{\href{https://orcid.org/0000-0002-3012-0316}{\includegraphics[scale=0.5]{Figures/orcid.jpg}}}$, 
S. G. Sousa\inst{10} $^{\href{https://orcid.org/0000-0001-9047-2965}{\includegraphics[scale=0.5]{Figures/orcid.jpg}}}$, 
Sz. Csizmadia\inst{2} $^{\href{https://orcid.org/0000-0001-6803-9698}{\includegraphics[scale=0.5]{Figures/orcid.jpg}}}$, 
A. Brandeker\inst{11} $^{\href{https://orcid.org/0000-0002-7201-7536}{\includegraphics[scale=0.5]{Figures/orcid.jpg}}}$, 
L. Carone\inst{4}, 
T. G. Wilson\inst{12} $^{\href{https://orcid.org/0000-0001-8749-1962}{\includegraphics[scale=0.5]{Figures/orcid.jpg}}}$, 
B. Akinsanmi\inst{5} $^{\href{https://orcid.org/0000-0001-6519-1598}{\includegraphics[scale=0.5]{Figures/orcid.jpg}}}$, 
J. A. Patel\inst{11}, 
A. Krenn\inst{4} $^{\href{https://orcid.org/0000-0003-3615-4725}{\includegraphics[scale=0.5]{Figures/orcid.jpg}}}$, 
O. D. S. Demangeon\inst{10,13} $^{\href{https://orcid.org/0000-0001-7918-0355}{\includegraphics[scale=0.5]{Figures/orcid.jpg}}}$, 
G. Bruno\inst{1} $^{\href{https://orcid.org/0000-0002-3288-0802}{\includegraphics[scale=0.5]{Figures/orcid.jpg}}}$, 
I. Pagano\inst{1} $^{\href{https://orcid.org/0000-0001-9573-4928}{\includegraphics[scale=0.5]{Figures/orcid.jpg}}}$, 
M. J. Hooton\inst{9} $^{\href{https://orcid.org/0000-0003-0030-332X}{\includegraphics[scale=0.5]{Figures/orcid.jpg}}}$, 
J. Cabrera\inst{2} $^{\href{https://orcid.org/0000-0001-6653-5487}{\includegraphics[scale=0.5]{Figures/orcid.jpg}}}$, 
N. C. Santos\inst{10,13} $^{\href{https://orcid.org/0000-0003-4422-2919}{\includegraphics[scale=0.5]{Figures/orcid.jpg}}}$, 
Y. Alibert\inst{7,6} $^{\href{https://orcid.org/0000-0002-4644-8818}{\includegraphics[scale=0.5]{Figures/orcid.jpg}}}$, 
R. Alonso\inst{14,15} $^{\href{https://orcid.org/0000-0001-8462-8126}{\includegraphics[scale=0.5]{Figures/orcid.jpg}}}$, 
J. Asquier\inst{16}, 
T. Bárczy\inst{17} $^{\href{https://orcid.org/0000-0002-7822-4413}{\includegraphics[scale=0.5]{Figures/orcid.jpg}}}$, 
D. Barrado Navascues\inst{18} $^{\href{https://orcid.org/0000-0002-5971-9242}{\includegraphics[scale=0.5]{Figures/orcid.jpg}}}$, 
S. C. C. Barros\inst{10,13} $^{\href{https://orcid.org/0000-0003-2434-3625}{\includegraphics[scale=0.5]{Figures/orcid.jpg}}}$, 
W. Baumjohann\inst{4} $^{\href{https://orcid.org/0000-0001-6271-0110}{\includegraphics[scale=0.5]{Figures/orcid.jpg}}}$, 
M. Beck\inst{5} $^{\href{https://orcid.org/0000-0003-3926-0275}{\includegraphics[scale=0.5]{Figures/orcid.jpg}}}$, 
T. Beck\inst{6}, 
W. Benz\inst{6,7} $^{\href{https://orcid.org/0000-0001-7896-6479}{\includegraphics[scale=0.5]{Figures/orcid.jpg}}}$, 
M. Bergomi\inst{19}, 
A. Bonfanti\inst{4} $^{\href{https://orcid.org/0000-0002-1916-5935}{\includegraphics[scale=0.5]{Figures/orcid.jpg}}}$, 
X. Bonfils\inst{20} $^{\href{https://orcid.org/0000-0001-9003-8894}{\includegraphics[scale=0.5]{Figures/orcid.jpg}}}$, 
L. Borsato\inst{21} $^{\href{https://orcid.org/0000-0003-0066-9268}{\includegraphics[scale=0.5]{Figures/orcid.jpg}}}$, 
C. Broeg\inst{6,7} $^{\href{https://orcid.org/0000-0001-5132-2614}{\includegraphics[scale=0.5]{Figures/orcid.jpg}}}$, 
S. Charnoz\inst{22} $^{\href{https://orcid.org/0000-0002-7442-491X}{\includegraphics[scale=0.5]{Figures/orcid.jpg}}}$, 
A. Collier Cameron\inst{23} $^{\href{https://orcid.org/0000-0002-8863-7828}{\includegraphics[scale=0.5]{Figures/orcid.jpg}}}$, 
M. B. Davies\inst{24} $^{\href{https://orcid.org/0000-0001-6080-1190}{\includegraphics[scale=0.5]{Figures/orcid.jpg}}}$, 
M. Deleuil\inst{25} $^{\href{https://orcid.org/0000-0001-6036-0225}{\includegraphics[scale=0.5]{Figures/orcid.jpg}}}$, 
A. Deline\inst{5}, 
L. Delrez\inst{26,27} $^{\href{https://orcid.org/0000-0001-6108-4808}{\includegraphics[scale=0.5]{Figures/orcid.jpg}}}$, 
B.-O. Demory\inst{7,6} $^{\href{https://orcid.org/0000-0002-9355-5165}{\includegraphics[scale=0.5]{Figures/orcid.jpg}}}$, 
D. Ehrenreich\inst{5,28} $^{\href{https://orcid.org/0000-0001-9704-5405}{\includegraphics[scale=0.5]{Figures/orcid.jpg}}}$, 
A. Erikson\inst{2}, 
L. Fossati\inst{4} $^{\href{https://orcid.org/0000-0003-4426-9530}{\includegraphics[scale=0.5]{Figures/orcid.jpg}}}$, 
M. Fridlund\inst{29,30} $^{\href{https://orcid.org/0000-0002-0855-8426}{\includegraphics[scale=0.5]{Figures/orcid.jpg}}}$, 
D. Gandolfi\inst{31} $^{\href{https://orcid.org/0000-0001-8627-9628}{\includegraphics[scale=0.5]{Figures/orcid.jpg}}}$, 
M. Gillon\inst{26} $^{\href{https://orcid.org/0000-0003-1462-7739}{\includegraphics[scale=0.5]{Figures/orcid.jpg}}}$, 
M. Güdel\inst{32}, 
M. N. Günther\inst{16} $^{\href{https://orcid.org/0000-0002-3164-9086}{\includegraphics[scale=0.5]{Figures/orcid.jpg}}}$, 
J.-V. Harre\inst{2}, 
A. Heitzmann\inst{5}, 
Ch. Helling\inst{4,33}, 
S. Hoyer\inst{25} $^{\href{https://orcid.org/0000-0003-3477-2466}{\includegraphics[scale=0.5]{Figures/orcid.jpg}}}$, 
K. G. Isaak\inst{16} $^{\href{https://orcid.org/0000-0001-8585-1717}{\includegraphics[scale=0.5]{Figures/orcid.jpg}}}$, 
L. L. Kiss\inst{34,35}, 
K. W. F. Lam\inst{2} $^{\href{https://orcid.org/0000-0002-9910-6088}{\includegraphics[scale=0.5]{Figures/orcid.jpg}}}$, 
J. Laskar\inst{36} $^{\href{https://orcid.org/0000-0003-2634-789X}{\includegraphics[scale=0.5]{Figures/orcid.jpg}}}$, 
A. Lecavelier des Etangs\inst{37} $^{\href{https://orcid.org/0000-0002-5637-5253}{\includegraphics[scale=0.5]{Figures/orcid.jpg}}}$, 
D. Magrin\inst{21} $^{\href{https://orcid.org/0000-0003-0312-313X}{\includegraphics[scale=0.5]{Figures/orcid.jpg}}}$, 
P. F. L. Maxted\inst{38} $^{\href{https://orcid.org/0000-0003-3794-1317}{\includegraphics[scale=0.5]{Figures/orcid.jpg}}}$, 
H. Mischler\inst{39}, 
C. Mordasini\inst{6,7}, 
V. Nascimbeni\inst{21} $^{\href{https://orcid.org/0000-0001-9770-1214}{\includegraphics[scale=0.5]{Figures/orcid.jpg}}}$, 
G. Olofsson\inst{11} $^{\href{https://orcid.org/0000-0003-3747-7120}{\includegraphics[scale=0.5]{Figures/orcid.jpg}}}$, 
R. Ottensamer\inst{32}, 
E. Pallé\inst{14,15} $^{\href{https://orcid.org/0000-0003-0987-1593}{\includegraphics[scale=0.5]{Figures/orcid.jpg}}}$, 
G. Peter\inst{40} $^{\href{https://orcid.org/0000-0001-6101-2513}{\includegraphics[scale=0.5]{Figures/orcid.jpg}}}$, 
G. Piotto\inst{21,41} $^{\href{https://orcid.org/0000-0002-9937-6387}{\includegraphics[scale=0.5]{Figures/orcid.jpg}}}$, 
D. Pollacco\inst{12}, 
R. Ragazzoni\inst{21,41} $^{\href{https://orcid.org/0000-0002-7697-5555}{\includegraphics[scale=0.5]{Figures/orcid.jpg}}}$, 
N. Rando\inst{16}, 
H. Rauer\inst{2,42,43} $^{\href{https://orcid.org/0000-0002-6510-1828}{\includegraphics[scale=0.5]{Figures/orcid.jpg}}}$, 
I. Ribas\inst{44,45} $^{\href{https://orcid.org/0000-0002-6689-0312}{\includegraphics[scale=0.5]{Figures/orcid.jpg}}}$, 
S. Salmon\inst{5} $^{\href{https://orcid.org/0000-0002-1714-3513}{\includegraphics[scale=0.5]{Figures/orcid.jpg}}}$, 
D. Ségransan\inst{5} $^{\href{https://orcid.org/0000-0003-2355-8034}{\includegraphics[scale=0.5]{Figures/orcid.jpg}}}$, 
A. E. Simon\inst{6,7} $^{\href{https://orcid.org/0000-0001-9773-2600}{\includegraphics[scale=0.5]{Figures/orcid.jpg}}}$, 
M. Stalport\inst{27,26}, 
M. Steinberger\inst{4}, 
Gy. M. Szabó\inst{46,47} $^{\href{https://orcid.org/0000-0002-0606-7930}{\includegraphics[scale=0.5]{Figures/orcid.jpg}}}$, 
N. Thomas\inst{6}, 
S. Udry\inst{5} $^{\href{https://orcid.org/0000-0001-7576-6236}{\includegraphics[scale=0.5]{Figures/orcid.jpg}}}$, 
B. Ulmer\inst{40}, 
V. Van Grootel\inst{27} $^{\href{https://orcid.org/0000-0003-2144-4316}{\includegraphics[scale=0.5]{Figures/orcid.jpg}}}$, 
J. Venturini\inst{5} $^{\href{https://orcid.org/0000-0001-9527-2903}{\includegraphics[scale=0.5]{Figures/orcid.jpg}}}$, 
E. Villaver\inst{14,15}, 
N. A. Walton\inst{48} $^{\href{https://orcid.org/0000-0003-3983-8778}{\includegraphics[scale=0.5]{Figures/orcid.jpg}}}$, 
T. Zingales\inst{49,21} $^{\href{https://orcid.org/0000-0001-6880-5356}{\includegraphics[scale=0.5]{Figures/orcid.jpg}}}$
}

\authorrunning{Singh et al.}
\institute{
\label{inst:1} INAF, Osservatorio Astrofisico di Catania, Via S. Sofia 78, 95123 Catania, Italy \and
\label{inst:2} Institute of Planetary Research, German Aerospace Center (DLR), Rutherfordstrasse 2, 12489 Berlin, Germany \and
\label{inst:3} INAF, Osservatorio Astrofisico di Torino, Via Osservatorio, 20, I-10025 Pino Torinese To, Italy \and
\label{inst:4} Space Research Institute, Austrian Academy of Sciences, Schmiedlstrasse 6, A-8042 Graz, Austria \and
\label{inst:5} Observatoire astronomique de l'Université de Genève, Chemin Pegasi 51, 1290 Versoix, Switzerland \and
\label{inst:6} Weltraumforschung und Planetologie, Physikalisches Institut, University of Bern, Gesellschaftsstrasse 6, 3012 Bern, Switzerland \and
\label{inst:7} Center for Space and Habitability, University of Bern, Gesellschaftsstrasse 6, 3012 Bern, Switzerland \and
\label{inst:8} ETH Zurich, Department of Physics, Wolfgang-Pauli-Strasse 2, CH-8093 Zurich, Switzerland \and
\label{inst:9} Cavendish Laboratory, JJ Thomson Avenue, Cambridge CB3 0HE, UK \and
\label{inst:10} Instituto de Astrofisica e Ciencias do Espaco, Universidade do Porto, CAUP, Rua das Estrelas, 4150-762 Porto, Portugal \and
\label{inst:11} Department of Astronomy, Stockholm University, AlbaNova University Center, 10691 Stockholm, Sweden \and
\label{inst:12} Department of Physics, University of Warwick, Gibbet Hill Road, Coventry CV4 7AL, United Kingdom \and
\label{inst:13} Departamento de Fisica e Astronomia, Faculdade de Ciencias, Universidade do Porto, Rua do Campo Alegre, 4169-007 Porto, Portugal \and
\label{inst:14} Instituto de Astrofisica de Canarias, Via Lactea s/n, 38200 La Laguna, Tenerife, Spain \and
\label{inst:15} Departamento de Astrofisica, Universidad de La Laguna, Astrofísico Francisco Sanchez s/n, 38206 La Laguna, Tenerife, Spain \and
\label{inst:16} European Space Agency (ESA), European Space Research and Technology Centre (ESTEC), Keplerlaan 1, 2201 AZ Noordwijk, The Netherlands \and
\label{inst:17} Admatis, 5. Kandó Kálmán Street, 3534 Miskolc, Hungary \and
\label{inst:18} Depto. de Astrofisica, Centro de Astrobiologia (CSIC-INTA), ESAC campus, 28692 Villanueva de la Cañada (Madrid), Spain \and
\label{inst:19} INAF - Osservatorio Astronomico di Padova \and
\label{inst:20} Université Grenoble Alpes, CNRS, IPAG, 38000 Grenoble, France \and
\label{inst:21} INAF, Osservatorio Astronomico di Padova, Vicolo dell'Osservatorio 5, 35122 Padova, Italy \and
\label{inst:22} Université de Paris Cité, Institut de physique du globe de Paris, CNRS, 1 Rue Jussieu, F-75005 Paris, France \and
\label{inst:23} Centre for Exoplanet Science, SUPA School of Physics and Astronomy, University of St Andrews, North Haugh, St Andrews KY16 9SS, UK \and
\label{inst:24} Centre for Mathematical Sciences, Lund University, Box 118, 221 00 Lund, Sweden \and
\label{inst:25} Aix Marseille Univ, CNRS, CNES, LAM, 38 rue Frédéric Joliot-Curie, 13388 Marseille, France \and
\label{inst:26} Astrobiology Research Unit, Université de Liège, Allée du 6 Août 19C, B-4000 Liège, Belgium \and
\label{inst:27} Space sciences, Technologies and Astrophysics Research (STAR) Institute, Université de Liège, Allée du 6 Août 19C, 4000 Liège, Belgium \and
\label{inst:28} Centre Vie dans l’Univers, Faculté des sciences, Université de Genève, Quai Ernest-Ansermet 30, 1211 Genève 4, Switzerland \and
\label{inst:29} Leiden Observatory, University of Leiden, PO Box 9513, 2300 RA Leiden, The Netherlands \and
\label{inst:30} Department of Space, Earth and Environment, Chalmers University of Technology, Onsala Space Observatory, 439 92 Onsala, Sweden \and
\label{inst:31} Dipartimento di Fisica, Università degli Studi di Torino, via Pietro Giuria 1, I-10125, Torino, Italy \and
\label{inst:32} Department of Astrophysics, University of Vienna, Türkenschanzstrasse 17, 1180 Vienna, Austria \and
\label{inst:33} Institute for Theoretical Physics and Computational Physics, Graz University of Technology, Petersgasse 16, 8010 Graz, Austria \and
\label{inst:34} Konkoly Observatory, Research Centre for Astronomy and Earth Sciences, 1121 Budapest, Konkoly Thege Miklós út 15-17, Hungary \and
\label{inst:35} ELTE E\"otv\"os Lor\'and University, Institute of Physics, P\'azm\'any P\'eter s\'et\'any 1/A, 1117 Budapest, Hungary \and
\label{inst:36} IMCCE, UMR8028 CNRS, Observatoire de Paris, PSL Univ., Sorbonne Univ., 77 av. Denfert-Rochereau, 75014 Paris, France \and
\label{inst:37} Institut d'astrophysique de Paris, UMR7095 CNRS, Université Pierre \& Marie Curie, 98bis blvd. Arago, 75014 Paris, France \and
\label{inst:38} Astrophysics Group, Lennard Jones Building, Keele University, Staffordshire, ST5 5BG, United Kingdom \and
\label{inst:39} Physikalisches Institut, University of Bern, Gesellschaftsstrasse 6, 3012 Bern, Switzerland \and
\label{inst:40} Institute of Optical Sensor Systems, German Aerospace Center (DLR), Rutherfordstrasse 2, 12489 Berlin, Germany \and
\label{inst:41} Dipartimento di Fisica e Astronomia "Galileo Galilei", Università degli Studi di Padova, Vicolo dell'Osservatorio 3, 35122 Padova, Italy \and
\label{inst:42} Zentrum für Astronomie und Astrophysik, Technische Universität Berlin, Hardenbergstr. 36, D-10623 Berlin, Germany \and
\label{inst:43} Institut fuer Geologische Wissenschaften, Freie Universitaet Berlin, Maltheserstrasse 74-100,12249 Berlin, Germany \and
\label{inst:44} Institut de Ciencies de l'Espai (ICE, CSIC), Campus UAB, Can Magrans s/n, 08193 Bellaterra, Spain \and
\label{inst:45} Institut d’Estudis Espacials de Catalunya (IEEC), Gran Capità 2-4, 08034 Barcelona, Spain \and
\label{inst:46} ELTE E\"otv\"os Lor\'and University, Gothard Astrophysical Observatory, 9700 Szombathely, Szent Imre h. u. 112, Hungary \and
\label{inst:47} HUN-REN–ELTE Exoplanet Research Group, Szent Imre h. u. 112., Szombathely, H-9700, Hungary \and
\label{inst:48} Institute of Astronomy, University of Cambridge, Madingley Road, Cambridge, CB3 0HA, United Kingdom \and
\label{inst:49} Dipartimento di Fisica e Astronomia "Galileo Galilei", Universita degli Studi di Padova, Vicolo dell'Osservatorio 3, 35122 Padova, Italy
}
\date{Received 24 July 2023 / Accepted 31 October 2023}

\abstract
        {Occultations are windows of opportunity to indirectly peek into the dayside atmosphere of exoplanets. High-precision transit events provide information on the spin-orbit alignment of exoplanets around fast-rotating hosts.}        
        {We aim to precisely measure the planetary radius and geometric albedo of the ultra-hot Jupiter(UHJ) \keltb\ along with the spin-orbit alignment of the system.}
        {We obtained optical high-precision transits and occultations of KELT-20 b using CHEOPS observations in conjunction with simultaneous TESS observations. We interpreted the occultation measurements together with archival infrared observations to measure the planetary geometric albedo and dayside temperatures. We further used the host star's gravity-darkened nature to measure the system's obliquity.}
        {We present a time-averaged precise occultation depth of 82$\pm$6 ppm measured with seven \cheops\ visits and 131$^{+8}_{-7}$ ppm from the analysis of all available \tess\ photometry. Using these measurements, we precisely constrain the geometric albedo of \keltb\ to 0.26$\pm$0.04 and the brightness temperature of the dayside hemisphere to 2566$^{+77}_{-80}$~K. Assuming Lambertian scattering law, we constrain the Bond albedo to $0.36^{+0.04}_{-0.05}$ along with a minimal heat transfer to the night side ($\epsilon=0.14^{+0.13}_{-0.10}$). Furthermore, using five transit observations we provide stricter constraints of $3.9\pm1.1$ degrees on the sky-projected obliquity of the system.}
        {The aligned orbit of \keltb\ is in contrast to previous CHEOPS studies that have found strongly inclined orbits for planets orbiting other A-type stars. The comparably high planetary geometric albedo of \keltb\  corroborates a known trend of strongly irradiated planets being more reflective. Finally, we tentatively detect signs of temporal variability in the occultation depths, which might indicate variable cloud cover advecting onto the planetary day side.}


\titlerunning{CHEOPS occultation of \keltb}
\maketitle
%
\section{Introduction}

Occultations (or secondary eclipses) are windows of opportunity to study exoplanetary atmospheres. An occultation depth is a measure of the planet's brightness that manifests in a subtle decrease in the observed flux as it goes behind (is occulted by) the star. This brightness consists of two components: the stellar light reflected off the planet's visible hemisphere and its thermal emission. For ultra-hot Jupiters (UHJs) observed in optical passbands, the contribution of the planet's thermal emission to its brightness is comparable to the predominant reflection contribution. Consequently, the measured occultation depth is a degenerate combination of the two, making it difficult to disentangle their respective contributions. This degeneracy can be broken with additional information, such as occultation depth measurements in different passbands. 

\keltb\ \citep{Lund2017AJ} or MASCARA-2b \citep{Talens2018} was first  reported simultaneously in the two discovery papers from the KELT \citep{KELTSurvey2018} and MASCARA \citep{MASCARA2012SPIE} transit surveys, respectively. Both surveys use dedicated ground-based telescopes for exoplanet discoveries, particularly the brightest transiting exoplanet systems. \keltb\ is a transiting UHJ that orbits one of the hottest (\teff\ $\sim$9000\,K) fast-rotating (\vsini$\sim$116\,km\,s$^{-1}$) stars in a $\sim$3.5 day orbit. This leads to an equilibrium temperature of \teqm\ $\sim$2300\,K (see Table~\ref{tab: stellar parameters} for details on stellar parameters). 

By virtue of many follow-up observations in the form of high-resolution transmission spectroscopy, the atmosphere of \keltb\ has been well characterised with the detection of the following species: Na, H$\alpha$, Mg, Fe, Fe I, Fe II, Ca$^{+}$, Cr II, Si \citep{Casasays-Barris2018A&A, Casasayas-Barris2020A&A, Stangret2020A&A, Hoeijmakers2020, Nugroho2020MNRAS, Cont2022A&A}. The day side of \keltb\ is hotter than the theoretical equilibrium temperature because of the intense far-ultraviolet (FUV) and ultraviolet (UV) radiation received from its host A2 star, which heats up the upper layers of the atmosphere\textbf{,} leading to the expansion and abrasion of species between different layers \cite{Fu2022ApJ}. The FUV/UV radiation also facilitates the presence of a strong thermal inversion in the underlying layers, which is evident from the additional detection of neutral and atomic iron in the emission lines \citep{Yan2022A&A, Borsa2022A&A}. Consequently, the atmosphere is inflated enough to show strong absorption in the transmission spectrum, and the host star is sufficiently hot to strengthen the atomic/molecular spectral signatures in the emission spectrum leading to a plethora of species discoveries, including the likes of CO and H2O \citep{Fu2022ApJ}. Furthermore, \citet{Rainer2021} observed both the classical and the atmospheric Rossiter–McLaughlin (RM) effect by studying the Fe\,I feature in transmission. This detection of the blue-shift in the atmospheric signal during the transit along with the well-resolved double peak of the Fe\,I spectral feature in the emission spectrum \citep{Nugroho2020MNRAS} provides possible indications of atmospheric variability.
\begin{table*}
\caption{Stellar parameters from SWEET-CAT catalogue used in this work.}             
\label{tab: stellar parameters}      
\centering          
\begin{tabular}{l c c c r}    
\hline\hline       
                     
Parameter & Symbol & Units & Value & Source\\
\hline

 &  &  & MASCARA-2 & \\
Aliases &  &  & HD\,185603 & Simbad \tablefootmark{1}\\
 &  &  &  TIC\,69679391 / TOI-1151 & \\
\hline
\cr
Spectral Type & & & A2V  & \citet{Talens2018}\\

Effective temperature & \teff & K & $8980^{+30}_{-130}$ & \citet{Lund2017AJ}\\
Surface gravity & $\log g$ & $\log{\mathrm{cm/s^2}}$ & $4.31\pm0.02$ & \citet{Lund2017AJ}\\

Density & $\rho_{\star}$ & $\mathrm{g/cm^3}$ &  $0.641^{+0.035}_{-0.033}$ & \citet{Lund2017AJ}\\

Metallicity & [Fe/H] & --- & $-0.02\pm0.07$ & \citet{Lund2017AJ}\\
Projected rotational velocity & \vsini & km/s & $116.23\pm1.25$ & \citet{Rainer2021} \\
Stellar rotation period & $P_{\mathrm{rot}}$ & d & 0.696$\pm$0.027 & \citet{Lund2017AJ} \\
RV semi amplitude &  & m/s & ${<322.51}$ &  \citet{Casasayas-Barris2019} \\

Stellar mass & $\rm M_\star$ & $\rm M_\sun$ & $1.89_{-0.05}^{+0.06}$ & \citet{Talens2018}\\
Stellar radius & $\rm R_\star$ & $\rm R_\sun$ & $1.60\pm0.06$ & \citet{Talens2018}\\

\hline                  
\end{tabular}
\tablefoot{
\tablefoottext{1}{SIMBAD astronomical database from the Centre de Donn\'ees astronomiques de Strasbourg.\footnote{\url{http://simbad.u-strasbg.fr/simbad/}}}
}
\end{table*}

The study of exoplanet atmospheres, particularly hot Jupiters, has already benefited from the high-precision occultation measurements with \cheops\ \citep{Lendl2020, Hooton2022A&A, Deline2022A&A, Brandekar2022A&A, Scandariato2022A&A, Jones2022A&A, Hannu2022A&A, Krenn2023A&A}. One of the new frontiers in exoplanetology is the detection of atmospheric variability measured as variations in the occultation depth and the phase curve's peak offset over time. Previous studies of atmospheric variability with \textit{Spitzer} in the canonical hot Jupiters HD 209458b and HD 189733b did not show any variability with robust statistical significance \citep[]{Kilpatrick12020AJ}. The first claim of atmospheric variability was announced for HAT-P-7~b \citep[]{Armstrong2016NatAs} using the \textit{Kepler} phase curves and later in Kepler-76~b \citep[]{Jackson2019AJ}. Both of these studies found strong variations in the longitude of the brightest point in the planets' atmospheres, although the former was recently attributed to photometric variability due to super-granulation in the host star \citep[]{Lally2022AJ}. With increasing photometric precision and multi-epoch observations, more reports of exoplanet variability become prevalent \citep[]{Hooton2019MNRAS, Wilson2021MNRAS, Hannu2022A&A}.

In this paper we present precise transit and occultation depths of \keltb\ using the data from its observation with the CHaracterising ExOplanet Satellite (\cheops\, \citealt{Benz2021}) and the Transiting Exoplanet Survey Satellite (\tess\, \citealt{Ricker2014SPIE}). We present all the datasets in Sect. \ref{sec:CHEOPSobs} together with the description of the data reduction. In Sect. \ref{sec: analysis} we describe the planetary light curve fitting algorithms used in this article. In Sect. \ref{sec: results} we present the results of our analysis, and discuss the outcomes in Sect. \ref{sec: discussion}. Finally in Sect. \ref{sec: conclusions} we draw our conclusions.

\section{Observations and data reduction}\label{sec:observations}
\subsection{\normalfont{\cheops} observations}\label{sec:CHEOPSobs}

CHEOPS observed the \kelt\ system in the consecutive summers of 2021 and 2022. The first set of observations include four secondary eclipses and four primary transits of the planet \keltb\ extended over a period of two months starting from June 25 to September 1, 2021, while the second includes three additional secondary eclipses and one more primary transit that were observed from July 2 to August 9,  2022 (\ref{fig:bestfits_cheops}). Each visit was centred around either the secondary eclipse or the primary transit using the ephemerides from \citet{Lund2017AJ}, with the total observation time equal to three times the eclipse or transit duration. Consequently, each visit spanned an average of 10 to 11~hr where a time series was acquired with a cadence of 36 s. Due to interruptions from Earth occultations and crossings over the South Atlantic Anomaly (SAA) and to data flagging by the data reduction pipeline (DRP), the efficiency (i.e. the fraction of the visit time leading to scientifically valuable observations) was   $\sim$67\%. All the information about the individual observations, which were part of two different CHEOPS Guaranteed Time Observation (GTO) subprogrammes, is summarised in Table \ref{tab:obs}.
\newline

\begin{table*}
\caption{Logbook of the \cheops\ observations of \kelt. The top seven are the occultation observations while the bottom five are the transits.}

\label{tab:obs}      
\centering          
\begin{tabular}{c c c c c c c}     
\hline\hline       
                      
File-key\tablefootmark{1} & Visit & Start time & Duration & N. frames\tablefootmark{2} & Efficiency\tablefootmark{3}& Rjct. pts. \\
& ID & {[UTC]} & [hr]  &  &  [\%] &  \\
\hline        

CH\_PR100016\_TG014101\_V0200  & 1520194 & 2021-06-25 12:43:51  & 10.98   &  747  &  69.6 & 78 \\
CH\_PR100016\_TG014102\_V0200  & 1533914 & 2021-07-12 22:48:52  & 11.22   &  695  &  64.2 & 70\\
CH\_PR100016\_TG014103\_V0200  & 1543424 & 2021-07-29 07:02:51  & 10.96   &  747  &  69.7 & 69\\
CH\_PR100016\_TG014104\_V0200  & 1566969 & 2021-08-12 04:01:51  & 10.98   &  772  &  71.5 & 80\\
CH\_PR100016\_TG014105\_V0200  & 1840687 & 2022-07-02 05:59:52  & 11.22    &  682  &  62.8 & 78\\
CH\_PR100016\_TG014106\_V0200  & 1850916 & 2022-07-16 03:59:51  & 10.87  &  705  &  66.6 & 74\\
CH\_PR100016\_TG014107\_V0200  & 1878526 & 2022-08-09 11:37:51  & 10.68  &  684  &  66.2 & 71\\
\hline
CH\_PR110047\_TG000401\_V0200  & 1532996 & 2021-07-11 04:05:52  & 11.0    &  664 &  62.8 & 64\\
CH\_PR110047\_TG000402\_V0200  & 1532997 & 2021-07-14 16:57:51  & 9.8     &  621  &  64.8 & 53\\
CH\_PR110047\_TG001001\_V0200  & 1564166 & 2021-08-11 10:50:51  & 9.31     &  612  &  67.7 & 63\\
CH\_PR110047\_TG001002\_V0200  & 1580776 &  2021-09-01 09:39:51  & 9.28   &  609  &  67.1 & 72\\
CH\_PR1\textbf{0}0020\_TG000801\_V0200  & 1869414 & 2022-07-28 08:48:49  & 10.83     &  716  &  64.9 & 79\\

\hline                  
\end{tabular}
\tablefoot{
\tablefoottext{1}{The file-key is the unique identifier associated with each CHEOPS visit.}
\tablefoottext{2}{A single frame is here the result of 3 stacked images and the integration time is $t_\text{int}=36\,\text{s}$ which is same for each observation. The last column indicates the number of the rejected data points.}
}
\end{table*}

\subsection{Data reduction pipeline}
Each CHEOPS observation was reduced by   version 13.1.0 of the CHEOPS Data reduction pipeline \citet[DRP, ][]{Hoyer2020}. The processing summary of the pipeline includes standard calibration steps such as bias, gain, non-linearity, dark-current, and flat fielding, and the correction for the environmental affects such as cosmic rays, smearing trails from nearby stars, and the sky background. Following the corrections, the photometric extraction was performed over four aperture sizes, three predefined and one optimised aperture size corresponding to the best signal-to-noise ratio (S/N). The default S/N (25 pixels)  is comparable to that of the optimal aperture size that varies between  15 and 19 pixels for different observations. The dispersion of the photometric light curve is lower with the former, and therefore we used the photometric extraction done with the default aperture radius instead of the optimal one. The field of view (FoV) of \kelt\ is populated with a few nearby background stars. The DRP estimated the contamination of these nearby objects by simulating the \cheops\ FoV based on the Gaia DR2 star catalogue \citep{GaiaCollaboration2020} and a template of the extended \cheops\ point spread function (PSF). To further analyse the DRP data, we used the PyCHEOPS software package developed specifically for the analysis of CHEOPS light curves \citep{Maxted2022MNRAS}.

\begin{figure*}
    \centering
    \includegraphics[width=.49\hsize,trim={0 0.5cm 2cm 1.5cm},clip]{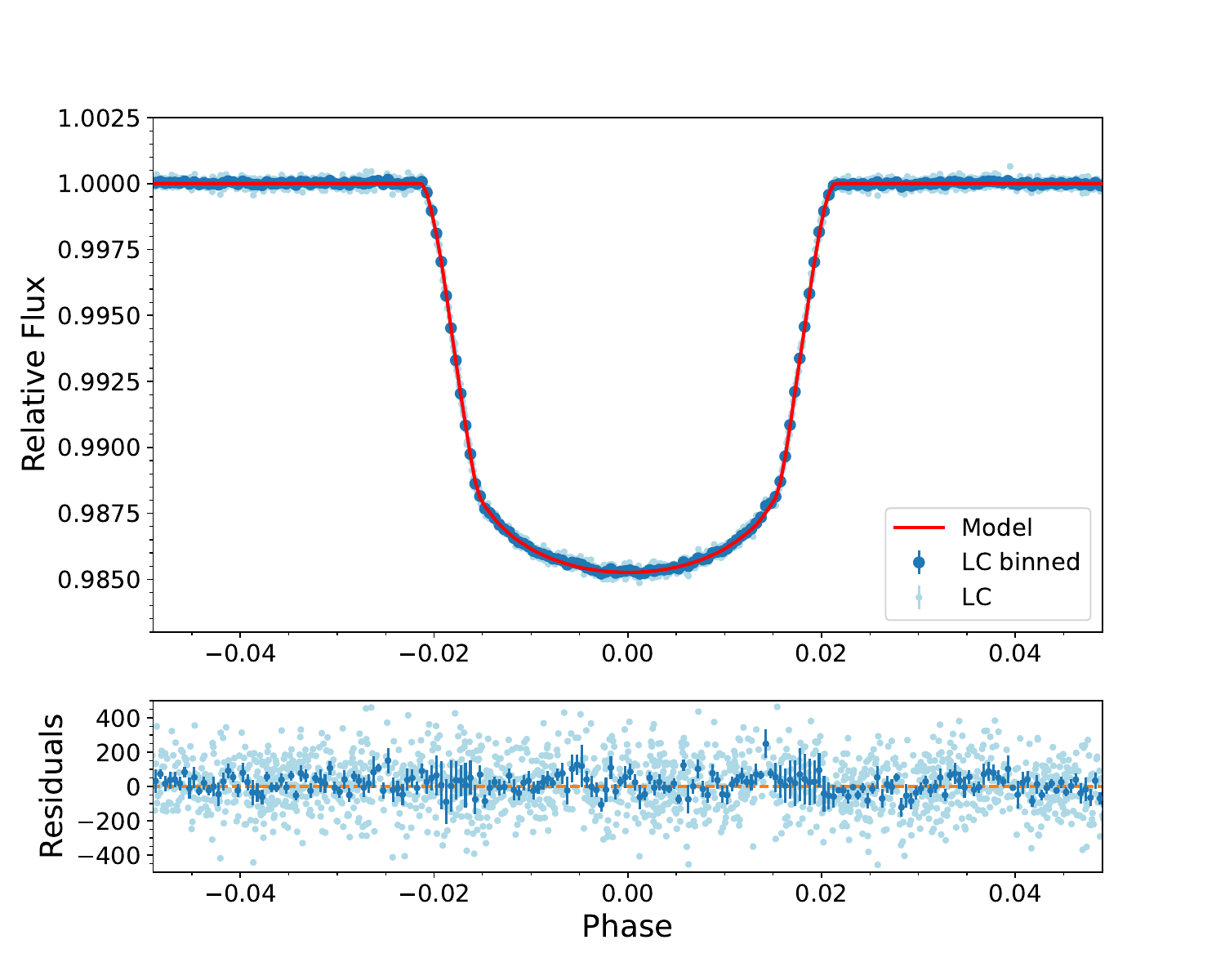}
    \includegraphics[width=.49\hsize,trim={0 0.5cm 2cm 1.5cm},clip]{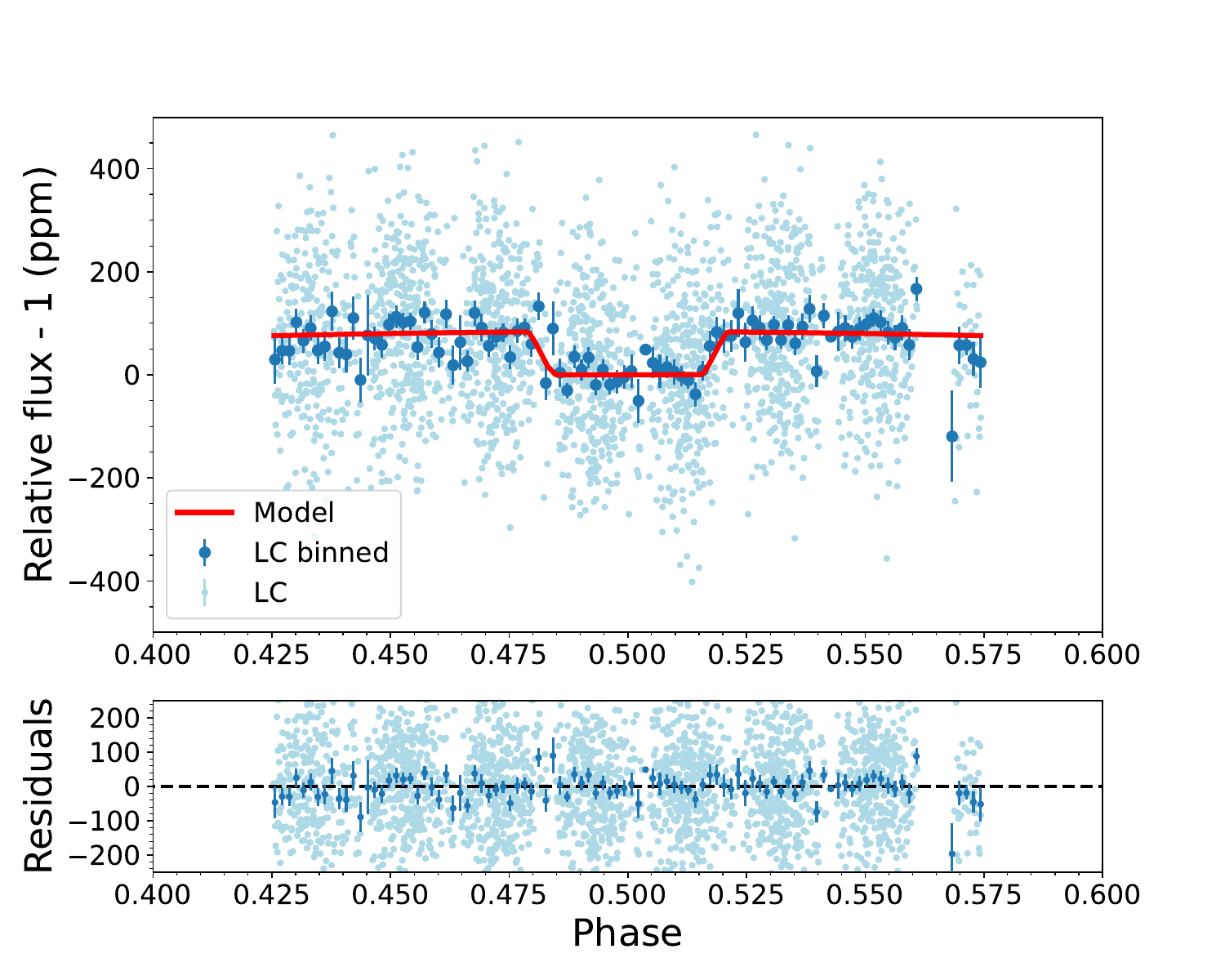}
    \caption{\cheops\ light curve. \textbf{\textit{Left:}} Detrended and phase-folded planetary transits. The solid red line is the best-fit model, while the blue dots represent the phase-folded and binned photometry (Interval: 2.25 minutes for transit, 7.5 minutes for occultation). The residuals are shown in the bottom panel. \textbf{\textit{Right:}} Same as in the left panel, but centred at the occultations.} \label{fig:bestfit_transit_occultation}
\end{figure*}  

\begin{figure*}
    \centering
    \includegraphics[width=.49\hsize,trim={0 0.5cm 2cm 1.5cm},clip]{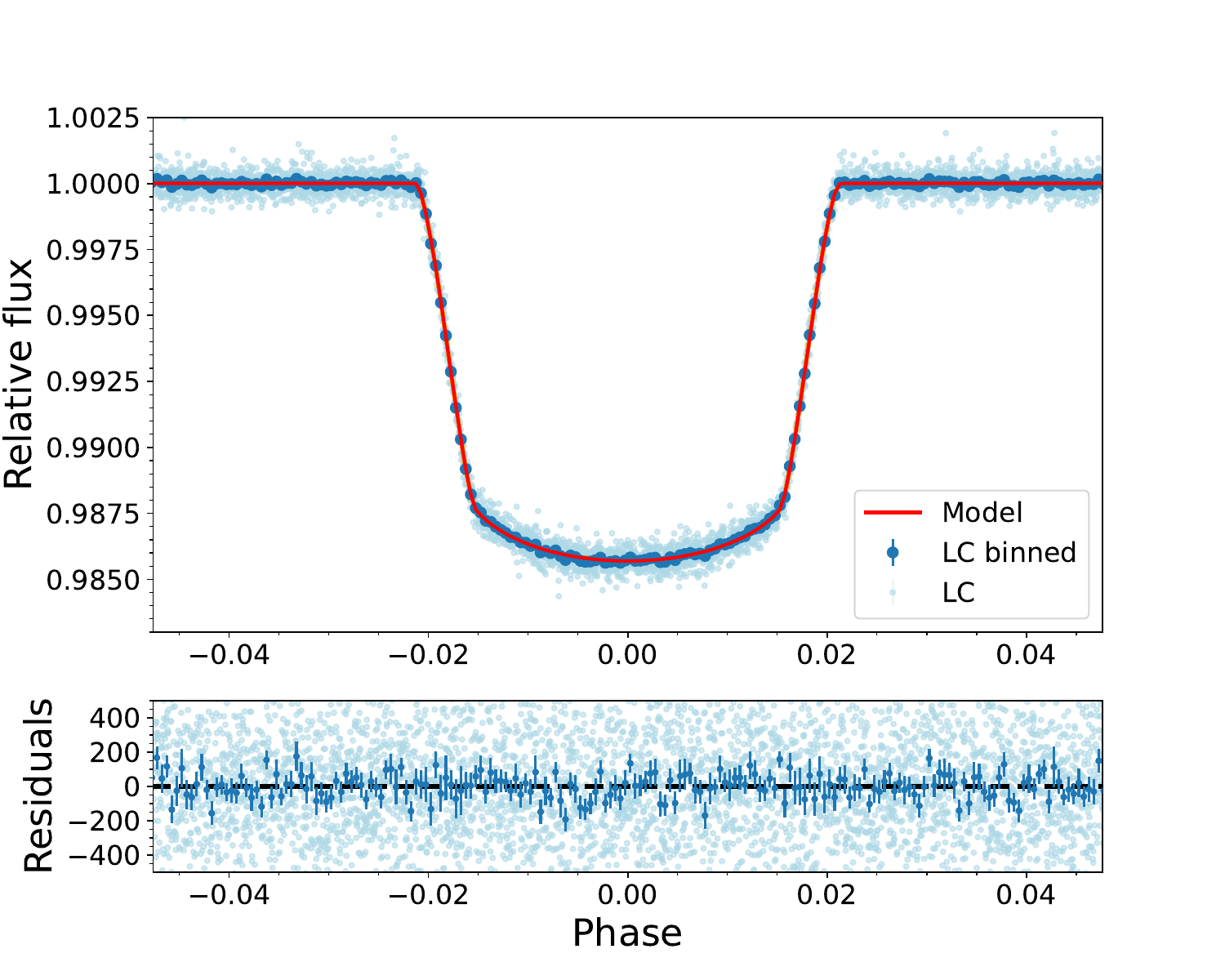}
    \includegraphics[width=.49\hsize,trim={0 0.5cm 2cm 1.5cm},clip]{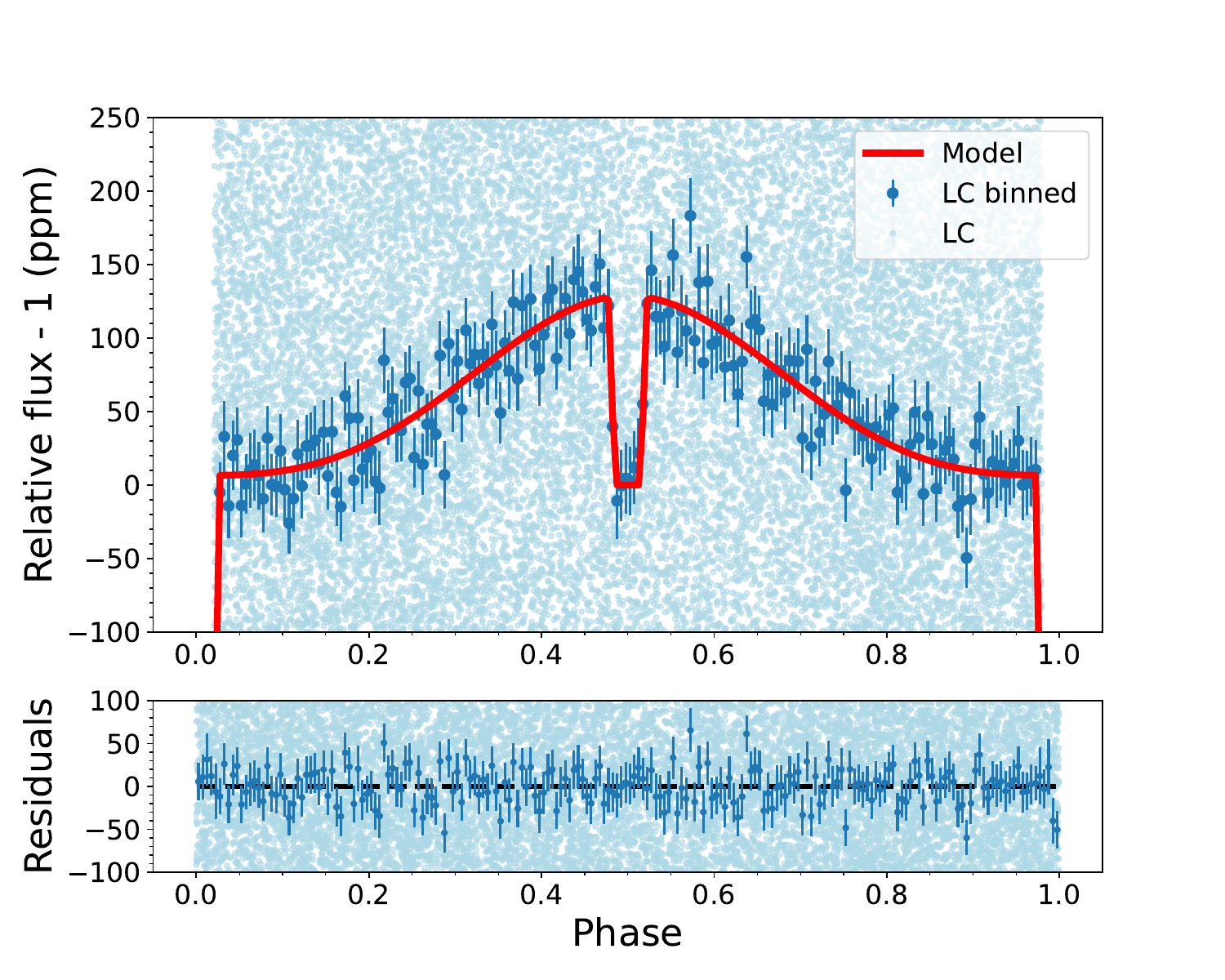}
    \caption{\tess\ light curve. \textbf{\textit{Left:}} Detrended and phase-folded planetary transits from Sectors 14, 40, 41, and 54. The solid red line is the best-fit model, while the blue dots represent the phase-folded and binned photometry (Interval: 2.25 minutes for transit, 25 minutes for phase curve). The residuals are shown in the bottom panel. \textbf{\textit{Right:}} Combined phase curve using photometry from all  four sectors.} \label{fig:bestfit_transit_eclipse_TESS}
\end{figure*}  
    
\subsection{Detrending using point spread function}
\label{sec:PSF detrending}

Due to the nature of the \cheops\ orbit, the presence of nearby background stars within the FoV can cause repeating flux variations on the timescales of the spacecraft's orbital period. These variations are in correlation with the roll angle of the telescope. Therefore, to account for such systematic changes, we need to simultaneously use sinusoidal detrending vectors to the DRP light curve during model fitting. These features are already present in the PyCHEOPS package, and we used them in our DRP light curve analysis. Additionally, it has been observed that such flux variations lead to subtle modifications to the target PSF (e.g. \citealt{Lendl2020,Bonfanti2021,Wilson2022}). Upon closer inspection of our \kelt\ data, we found this characteristic systematic, and thus applied the PSF-SCALPELS algorithm \citep{Wilson2022} to the DRP photometry. In brief, PSF-SCALPELS corrects flux modulations due to PSF shape changes by conducting a principal component analysis on the auto-correlation function of the photometric frames. This set of principal components traces variations in the PSF from the images, and is used in a linear noise model to detrend light curve fluxes. A more detailed explanation can be found in \citet{Wilson2022} with further examples of use shown in \citet{Hoyer2022}, \citet{Hannu2022A&A}, \citet{Hawthorn2023}, and \citet{Ehrenreich2023}. 
\par
While PSF-SCALPELS is a PSF-based correction to the DRP photometry, we also performed an independent photometric extraction using the PSF Imagette Photometric Extraction pipeline (PIPE) package\footnote{\url{https://github.com/alphapsa/PIPE}} \cite{Morris55cnce, Brandekar2022A&A}. We found that the dispersion in the reduced light curves are compatible with that of the other extraction techniques. They vary around a median absolute deviation (MAD) of 150 ppm at the raw cadence of 36 seconds. The light curves with PIPE extraction are free from roll angle variations and do not require additional detrending. Consequently, the error bars on the resultant fit parameters are smaller compared to those obtained via other techniques. We therefore chose to use the PIPE extracted photometry for further light curve analysis. Nevertheless, we show a comparative measurement of the occultation depths analysed using the three data extraction pipelines in Fig. \ref{fig: ComparativeVariability}. This exercise demonstrates the robustness of the extraction techniques, and further enhances confidence in the final outcome.

\subsection{Outlier removal}
Prior to the light curve analysis, we first filtered out the generic outliers flagged by PIPE that could have originated from cosmic ray encounters, centroid shifts far away from the median centroid, noisy bad pixels, or a poor PSF fit.\textbf{\footnote{https://pipe-cheops.readthedocs.io/en/latest/index.html}} The data points in the light curve corresponding to these images were masked out from further analysis. In addition, we also discarded data points where the flux is far above or below the median flux by setting a limit of five times the MAD value.

During each \cheops\ orbit, the observations are interrupted by Earth occultations. In proximity to the occultations, the background flux is drastically pushed up by scattered light from the Earth, leading to increased scatter in the light curve as it closes to the Earth's occultation. We decided to remove the data points whose corresponding background flux exceeds the median value by more than three times the median absolute deviation. We proceeded with the leftover good data points and fitted them with the planetary model discussed in the next section.

\subsection{\normalfont{\tess} observations}\label{sec:TESSobs}

TESS observed the \kelt\ system in Sector 14 (July 18 to August 14, 2019), Sectors 40--41 (June 25 to August 20, 2021), and in Sector 54 (July 9 to August 5, 2022). We downloaded the 2 min cadence pre-searched data conditioning single aperture photometry (PDC-SAP) light curves \citep{Jenkins2016} using the \texttt{lightkurve} package \citep{lightKurve2018}. The light curves were corrected for instrumental systematics following \citet{Smith2012, Stumpe2012, Stumpe2014}. We  also experimented with single aperture photometry (SAP) light curves. We found conflicting results between the PDC-SAP and the SAP detrended light curve analysis;  the results of the former are consistent with the CHEOPS results, and therefore we present only the PDC-SAP light curve analysis.

\section{Light curve analysis}\label{sec: analysis}
\subsection{Planetary model}\label{sec: Planetary Model}
Our light curve model $F(\phi)$ is defined as the sum of the contributions from several independent submodels related to planetary and stellar flux variations. These include a primary transit $F_{\rm tr}(\phi)$ and an occultation $F_{\rm occ}(\phi)$ model with a flat out-of-eclipse baseline; a phase variation $F_{\rm day}(\phi)$ model corresponding to the flux from the planetary day side and a similar nightside phase variation term $F_{\rm nt}(\phi)$, an ellipsoidal variation \citep{Morris1985ApJ, Mislis2012A&A} term $F_{\rm EV}(\phi)$, and a Doppler boosting \citep{LoebGaudi2003ApJ, BarbierLopez2021RMxAA} term $F_{\rm DB}(\phi)$ corresponding to the changes in stellar flux. For \keltb\ the combined amplitude of $F_{\rm EV}(\phi)$ and $F_{\rm DB}(\phi)$ are of the order of unity ($\sim$2 ppm), which stands below the measured uncertainties, while the phase amplitude corresponding to the planetary brightness is of the order of tens of ppm, and therefore we excluded the former from our phase curve modelling. The final model constitutes the following:
\begin{equation}\label{eqn: LC model}
    F(\phi) = f_{0} + F_{\mathrm{tr}}(\phi) + F_{\mathrm{occ}}(\phi) + F_{\mathrm{day}}(\phi) + F_{\mathrm{night}}(\phi), 
\end{equation}

\noindent 
where $\phi$ is the planet's orbital phase and $f_{0}$ is an arbitrary additive offset to align the light curve such that the base of the eclipse is at a relative flux value equal to one. The complete mathematical description can be found in \citet{Esteves2013ApJ} and references therein. Our model assumes a tidally locked circular orbit for \keltb\ following \citet{Wong2021}. We used the \citet{MandelAgol2002} transit model of a circular disc passing over a spherical stellar disc with the following parametrisation: 
\begin{itemize}
    \item mid-transit epoch $T_{0}$;
    \item orbital frequency $\nu_{\mathrm{orb}}$ (inverse of the orbital period $P_{\mathrm{orb}}$);
    \item stellar density $\rho_{\star}$;
    \item planet to star radius ratio $R_{p}/R_{\star}$;
    \item impact parameter $b$;
    \item the re-parametrised quadratic limb darkening coefficients $q_1$ and $q_2$ \citep{Kipping2013MNRAS}.
\end{itemize}
The occultation model follows a similar description, with the difference that the smaller disc is passing behind the larger one and without any limb darkening law. This  description was applied in previous works \citep{Singh2022, Scandariato2022A&A}. The planetary brightness at superior conjunction is parametrised by an occultation depth $\delta_{\mathrm{occ}}$. The phase variation term $F_{\mathrm{day}}(\phi)$ models the changing planetary flux emanating from the day side during its orbital motion, which is parametrised by an amplitude term $A_{\mathrm{day}}$ and a brightness variation term defined by the reflected light phase function \citep{Esteves2015ApJ, Singh2022}. The dayside phase variation is a consequence of Lambertian radiation from both reflection and thermal emission. A dayside phase offset $\Delta\phi$ is included to account for the asymmetry in the phase variation \cite[]{Scandariato2022A&A}. We followed the convention where positive $\Delta\phi$ implies an eastward phase shift, while negative is for westward shift. We also incorporate a similar but inverted phase function for thermal emission from the night side, which is parametrised by $A_{\mathrm{night}}$. No phase offset term is included, and therefore it peaks at mid-transit, while the minima is at mid-eclipse. The minima of the dayside flux is at mid-transit, while the minima of the nightside flux is at mid-occultation. The eclipse depth from these three parameters is obtained by the following equation:
\begin{equation}
    \delta_{\mathrm{occ}} = F_{\mathrm{day}}(\phi = 0.5) + A_{\mathrm{night}}.
\end{equation}

We followed this description along with the transit parameters while fitting the full phase curve. Phase curves are only available for the \tess\ light curves;  the \cheops\ observations do not have phase information, and therefore we followed a simpler description with $F_{\mathrm{night}} = 0$. To take into account that white noise is not included in the formal photometric uncertainties, we added an independent jitter term to our model.

We fitted the data with the combined model and searched for the simultaneous convergence of the free parameters by maximising the log-likelihood function in the parameter space. We sampled the posterior probability distribution of the free model parameters in a Bayesian MCMC framework using the emcee package version 3.0.2 \citep{Foreman-Mackey2013}. We followed the base line criterion of convergence following the methods described in \citet{Goodman&Weare2010}. Under the situations where we had to choose among a set of different models, we followed the Akaike information criterion (AIC, \citet{burnham1998practical}) because of its robustness in information extraction without strongly penalising any additional model parameters. Given the complexity and demand of resources, we performed all our model fitting on the AMONRA computing infrastructure \citep{Bertocco2020, Taffoni2020}. 

\subsection{Transit model with gravity darkening}
\label{sec:grav_dark}
\kelt\ is an A2V star with a \vsini of 116 km/s and a radius of 1.6~$R_{\sun}$ resulting in a rotation period of $\sim$1 day and is therefore categorised as a fast-rotator. The centrifugal force at the surface of  fast-rotating stars is non-negligible  compared to the stellar gravitational force. Consequently, the stellar gravity is highest at the poles and lowest at the equators, creating an oblate star. According to the von Zeipel theorem, the radiative flux at a given latitude of a fast-rotating star is proportional to the local effective gravity \citep{vonZeipel1924MNRAS}. The flux radiated at equatorial stellar latitudes is significantly lower compared to the poles. This effect is referred to as gravity darkening (GD). During a transit, when a planet passes over and hides the varying brightness of the stellar surface, it generates an asymmetric transit curve, especially when the orbit is misaligned. The greater the projected orbital obliquity $\lambda_{p}$, the larger  the asymmetry \citep{Lendl2020, Deline2022A&A, Hooton2022A&A}. In addition, high-precision transit analysis is the  preferred technique in measuring the true obliquity ($\Psi$) of a planetary system. Using the stellar and orbital inclinations, it is computed by the following relation
\begin{equation}\label{eq:trueobliquity}
    \cos{\Psi}=\cos{i_\star}\cos{i_\mathrm{p}}+\sin{i_\star}\sin{i_\mathrm{p}}\cos{\lambda}.
\end{equation}

We model the GD transits of \kelt\ using the Transit Light Curve Modeler (TLCM; \citealt{TLCM}, \citealt{Csizmadia_wavelets1}), which was used   to model GD in the WASP-189 \citep{Lendl2020}, MASCARA-1 \citep{Hooton2022A&A}, and WASP-33 \citep{kalman2022_wasp33} systems. The fitted GD parameters in TLCM are the inclination of the stellar rotation axis, $i_\star$, and the longitude of the node of the stellar rotation axis, $\Omega_\star$, which is directly related to the sky-projected spin-orbit obliquity, $\lambda$, via $\cos \lambda = \pm \cos(\Omega_\mathrm{p} - \Omega_\star)$, where $\Omega_\mathrm{p}$ is the planetary rotational axis and is set to $90^\circ$. The GD exponent, $\beta$, is typically set to its theoretically expected value of 0.25 following von Zeipel’s law \citep{vonZeipel1924MNRAS}. Along with the GD parameters, the model describes the transits using the following fitted parameters: the scaled orbital semi-major axis, $a/R_\star$; the planet-to-star radius ratio, $R_\mathrm{p}/R_\star$; the transit impact parameter, $b$; the epoch of mid-transit; the orbital period, $P$; and the limb-darkening parameters, $u_+$ and $u_-$, related to the quadratic limb-darkening parameters, $u_1$ and $u_2$, via $u_+ = u_1 + u_2$ and $u_- = u_1 - u_2$. Two additional parameters, $\sigma_\mathrm{w}$ and $\sigma_\mathrm{r}$, describe respectively the levels of white and red (correlated) noise present in the light curves, determined using the wavelet formulation of \cite{Carter_Winn_wavelets}.


\section{Results}\label{sec: results}
In this section we present the outcomes of several of our analyses that include  seven occultation and five transit observations of \keltb\ with \cheops\ spanning over more than a year. Additionally, we also analysed the four \tess\ sectors of observations of \kelt\ that spanned over three years. We first present the results of the \cheops\ transit analysis with and without GD. We then present the results of only the occultation fits of both the telescopes followed by the results of the \tess\ phase curves.


\begin{table*}[h!]
\caption[]{Best-fit model and derived parameters of the transit, occultation, and phase curve fit. G.D. represents gravity-darkening fit.}
\label{tab:results}
\begin{tabular}{lccll}
    \hline \hline
    \multicolumn{2}{c}{Parameter} & Unit & CHEOPS & TESS\\ 
    \hline 
    \multicolumn{5}{l}{\textit{\textbf{Transit}}} \\
    \hline
    Stellar density & $\rho_{\star}$ & $\mathrm{g/cm^3}$ & $0.6501^{+0.0057}_{-0.0056}$ & $0.6444^{+0.0053}_{-0.0052}$\\ 
    Orbital frequency & $f$ & $\mathrm{d^{-1}}$ & $0.28784401\pm(33)$ & $0.28784433\pm(23)$\\ 
    Orbital Period & $P$ & d & $3.4741039\pm(40)$ & $3.47409999\pm(28)$ \\
    Orbital Period [G.D.] & - & d & $3.47410075\pm(56)$ & $3.4741004\pm(04)$ \\
    &&&&\\
    Time of transit & $T_{c}$ & BJD$_{\rm TDB}-2457000$ & 2406.927174$\pm$(24) & 1698.210775$\pm$(52)\\ 
    Time of transit [G.D.] & - & - & $1312.5853\pm(02)$ & $1312.5849\pm(01)$ \\ 
    &&&&\\
    Planet-star radius ratio & $R_{p}/R_{\star}$ & & $0.11572\pm0.00018$ & $0.11588\pm0.00006$ \\ 
    Planet-star radius ratio [G.D.] & $R_{p}/R_{\mathrm{eq}}$ & & 0.1131$\pm$0.0008 & $0.1121\substack{+0.0005\\-0.0003}$ \\ 
    &&&&\\
    Impact parameter & $b$ &  & $0.515\pm0.005$ & $0.523\pm0.004$ \\
    Impact parameter [G.D.] & - &  & 0.498$\pm$0.005 & $0.510\pm0.006$ \\ 
    &&&&\\
    Quadratic limb darkening & $q_\mathrm{1}$ &  & $0.257\pm0.018$ &  0.12$\pm$0.05 \\ 
    Quadratic limb darkening & $q_\mathrm{2}$ &  & $0.322^{+0.036}_{-0.034}$ & 0.39$\pm$0.26 \\
    Quadratic limb darkening [G.D.] & $u_{a} + u_{b}$   &   & 0.516$\pm$0.021 & $0.18\pm0.03$\\
    Quadratic limb darkening [G.D.] & $u_{a} - u_{b}$   &   & $0.05\pm0.10$ &  $0.39\pm0.02$ \\

    &&&&\\
    Scaled semi-major axis & $a/R_\star$ &  & $7.4579^{+0.0217}_{-0.016}$ & $7.4359^{+0.0204}_{-0.0203}$\\ 
    Scaled semi major axis [G.D.] & - & - & 7.46$\pm$0.02 & 7.43$\pm$0.02 \\
    &&&&\\
    Stellar rotation axis & $i_{\star}$ & deg &  88.9$^{+18.0}_{-19.8}$           & $55\substack{+9 \\ -5}$ \\
    Nodal longitude & $\Omega_{\star}$ & deg  &  86.12$\pm1.10$     & 81$\pm$1 \\
    Orbital inclination & $i_{\rm p}$ & deg & $86.03\pm0.05$ & $85.96\pm0.04$\\ 
    Orbital inclination [G.D.] & - & deg & $86.17\pm0.04$ & $86.064\pm0.047$\\ 
    &&&&\\
    Transit depth & $\delta_{\rm tr}$  & \% & $1.3391 \pm 0.0042$ & $1.3428\pm0.0013$ \\
    Transit depth [G.D.] & $\left(\frac{R_{p}}{\sqrt{R_{\mathrm{pol}}R_{\mathrm{eq}}}}\right)^{2}$ & \% & 1.3187 $\pm$ 0.019 & 1.296 $\pm$ 0.012 \\
    Sky-projected spin-orbit angle & $\lambda$ & deg & $3.9\pm1.1$  & $9\pm1$\\
    Spin-orbit angle & $\Psi$ & deg & $5.0\pm11$ & $32\pm7$\\ 
    &&&&\\
    Jitter & $\sigma_{\rm w}$  & ppm & $123 \pm 2$ & $303 \pm 4$ \\
    Red noise & $\sigma_{r}$ & ppm & $1947\pm107$ &   $7862\pm284$ \\
    &&&&\\

    \multicolumn{5}{l}{\textit{\textbf{Occultation and phase-curve}}}
    \\
    \hline
    Transit/Eclipse duration & T$_{14}$ / t$_{14}$ & hr & $3.540\pm0.003$ & $3.537\pm0.003$\\ 
    Phase offset & $\Delta\phi$ & deg & - & $-7^{+4}_{-7}$ \\ 
    
    Dayside Phase amplitude & $A_{\rm day}$  & ppm  & - & 122.2$\pm$4.9\tnote{*} \\
    Nightside Phase amplitude & $A_{\rm night}$  & ppm  & - & $8.6_{-5.8}^{+8.0}$ \\
    Occultation depth & $\delta_{\rm occ}$ & ppm & $82.4\pm5.6$ & $131^{+8}_{-7}$\\

    Brightness temperature (100\% NIR) & $T_{\rm day}$ & K & 2759$^{+200}_{-250}$ & 2555$^{+200}_{-250}$\\ 
    Geometric albedo (100\% NIR)& $A_{\rm g}$ &  & 0.19$\pm$0.08 & 0.19$\pm$0.16 \\

    \\
    \hline
    \multicolumn{5}{l}{\textit{Lambertian grey-sky (\cheops\ -- \tess)}}\\
    Brightness temperature & $T_{\rm day}$ & K &      \hspace{70pt}2566$^{+77}_{-80}$ & \hspace{-50pt} \\ 
    Geometric albedo & $A_{\rm g}$ &  & \hspace{70pt}$0.26 \pm 0.04$ & \hspace{-50pt} \\
    Dayside temperature & $T_\mathrm{d}$ & K & \hspace{70pt}$2618^{+35}_{-40}$& \hspace{-50pt}\\
    Nightside temperature & $T_\mathrm{n}$ & K & \hspace{70pt}$1266^{+259}_{-330}$ & \hspace{-50pt}\\ 
    Bond albedo & $A_{\rm B}$ &  & \hspace{70pt}$0.36^{+0.04}_{-0.05}$ & \hspace{-50pt}\\
    Heat re-circulation efficiency & $\epsilon$ &  & \hspace{70pt}$0.14^{+0.13}_{-0.10}$ & \hspace{-50pt}\\
    \hline 
\end{tabular}
\tablefoot{
\tablefoottext{1}$R_{\mathrm{eq}}$ and $R_{\mathrm{pol}}$ are the equatorial and the polar radius of the star, respectively. \\
\tablefoottext{2}$T_{14}$ = $t_{14}$: The transit and occultation duration are the same, as eccentricity is set to zero for a circular orbit.\\
\tablefoottext{3}$100\%$ NIR: 1D retrieval model without scattering.
}
\end{table*} 

\subsection{Transits}
\subsubsection{Without gravity darkening}
We fitted the five \cheops\ transit light curves individually following the model described in Sect.~\ref{sec: Planetary Model}, where all the component models except for the transit ($F_{\mathrm{tr}}$) are set to zero. The best-fit parameters of the individual transit fits are reported in Table~\ref{tab: transits_individual}, while the same for the combined fit of all the transits are reported in Table \ref{tab:results}. We present the best-fit transit model plotted over the phase folded photometry of all five transits in Fig.~\ref{fig:bestfit_transit_occultation} (on the left). We also present the individual transit fits in Fig. \ref{fig:bestfits_cheops}. All the transits appear symmetrical with converging limb darkening coefficients, thereby implying an equatorial orbit for the hot Jupiter around the fast-rotator host. We compare the best-fit limb darkening coefficients ($q_1$ and $q_2$) reported in Table \ref{tab:results} with the theoretical estimates of $q_1=0.250\pm0.001$ and $q_2=0.288\pm0.001$, computed using the  LDTk Python package \cite{HannuLDTK2015MNRAS},\footnote{https://github.com/hpparvi/ldtk} and found that their agreement is close to $1~\sigma$. This description is fairly consistent with that obtained with Doppler tomography \citep{HoeijmakersEtal2020aaWASP121bHiresSpecies}. The transit parameters observed by \cheops\ agree with the measurements of the \tess\ photometry described later in Sect.~\ref{sec: phase curves}. Figure~\ref{fig:bestfit_transit_eclipse_TESS} shows the best-fit plot of the \tess\ transits. The  outcome of the \tess\ photometry agrees with previously reported literature values \citep{Patel_Espinoza2022AJ, Wong2021}.

\subsubsection{With gravity darkening}
With the aim of better constraining the obliquity of the \kelt\ system, we fitted the CHEOPS transit light curves with a GD model as described in Sect. \ref{sec:grav_dark}. We placed a uniform prior  $\mathcal{U}(80^\circ,110^\circ)$ on $\Omega_\star$, corresponding to $-10^\circ < \lambda < 10^\circ$, based on Doppler tomography measurements of $\lambda$, which  indicate  that the \kelt\ system has an obliquity (sky-projected) close to zero \cite[]{Hoeijmakers2020}. Without this prior, the obliquity is poorly constrained by the transit photometry alone. We report our sky-projected ($\lambda$) and true obliquity ($\Psi$) values along with the other gravity darkening parameters in Table \ref{tab:results}. The measured sky-projected obliquity in \tess\ differs from the same measured with \cheops. This could be due to some unaccounted for noise sources or degeneracy in the fitting the \tess\  light curve. We present and compare our best-fit $\lambda$ along with the literature values in Section \ref{sec: spinorbit}. The flattening parameter (f) is 0.03, which gives the planet's polar-to-equatorial radius ratio $R_{\mathrm{pol}}/R_{\mathrm{eq}}$ at 0.97. Consequently, the planet-to-star radius ratio differs by more than 3$\sigma$ with and without GD model considerations. However, using the geometric mean as the stellar radius (i.e. $R_{\star} = \sqrt{R_{\mathrm{pol}}R_{\mathrm{eq}}}$ )  we reached an agreement between the two radius ratios.   

We performed a similar analysis over the combined transits of the \tess\ photometry, and present it  in Table \ref{tab:obliquity}. The fitting algorithm returns quite different outcomes   compared to the \cheops\ transit fits. It returns a significantly inconsistent orientation of the \kelt\ system with a much lower radius ratio. This could be primarily attributed to the lower precision of the \tess\ light curves, and therefore a non-GD model best describes the transit light curve. Each sector's individual \tess\ transits are shown in Figs.~\ref{fig:bestfit_transits_sector_14_40} and \ref{fig:bestfit_transits_sector_41_54}.

\subsection{Occultations}\label{sec: results_occ}
\begin{figure*}
  \begin{minipage}[c]{0.75\textwidth}
    \includegraphics[width=\textwidth]{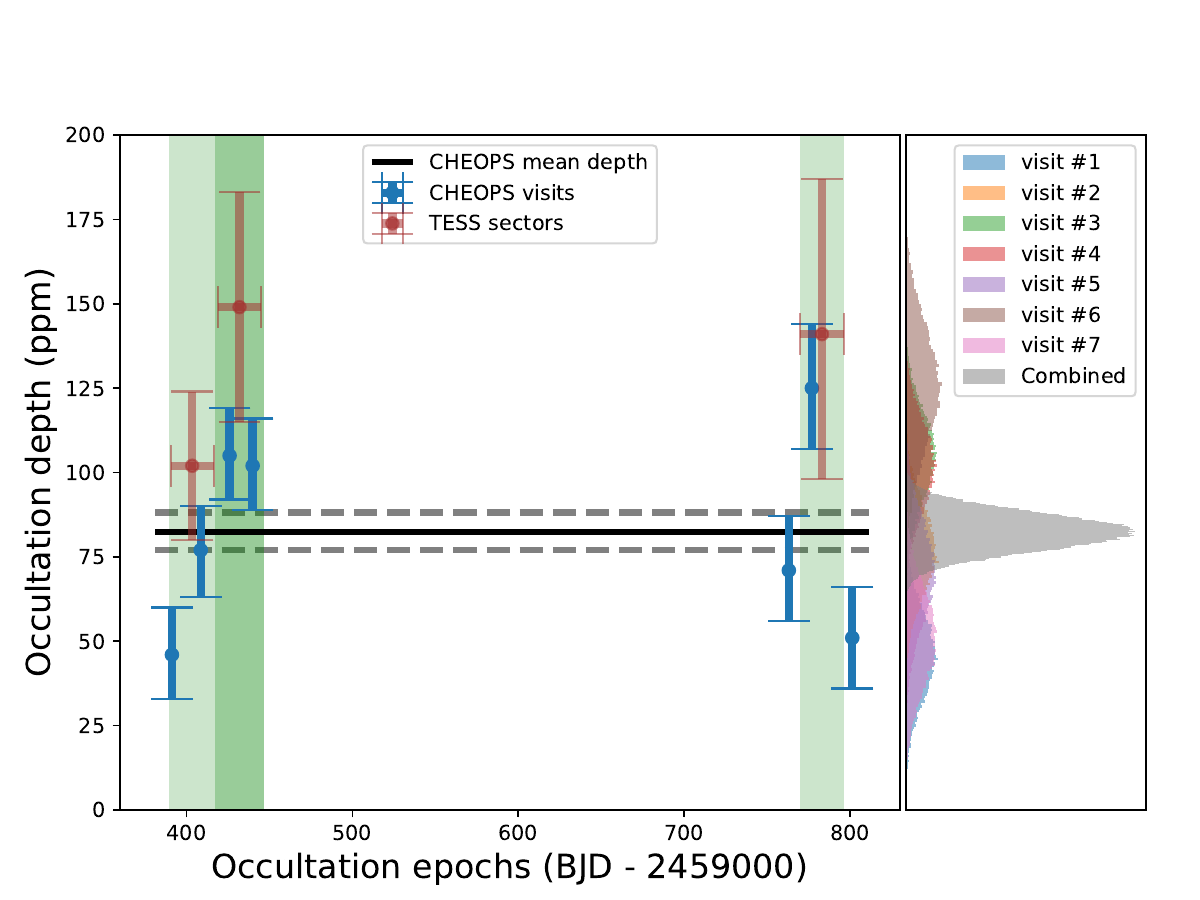}
  \end{minipage}\hfill
  \begin{minipage}[c]{0.25\textwidth}
    \caption{Best-fit occultation depths of individual \cheops\ visits (blue) along with their posteriors on the right. The horizontal solid and dashed lines represent the mean occultation depth of 82.4 $\pm$ 5.6 ppm. The panel on the  right  displays the best-fit posterior distribution. The green shaded regions indicate the span of the \tess\ Sectors 40, 41, and 54, while the brown points indicate the sector-wise occultation depths.} \label{fig:variability}
  \end{minipage}
\end{figure*}

CHEOPS observed a total of seven occultations of \keltb, which we analysed individually and jointly. For these analyses we fitted the model as defined in Eq.~\ref{eqn: LC model}, where the only free parameter is the occultation depth. We fixed the transit ephemerides following the \tess\ phase curve analysis described later in Sect.~\ref{sec: phase curves}. We first fitted the data of only the occultation model along with a jitter term. Following which we found a correlation with the time, x and y pixel centroids, and the roll angle, and therefore we decided to include them as a co-detrending term. We obtain a clear detection of the occultation depth in all the visits with a precision of more than 3\,$\sigma$ for the first and the last visit, and more than 5\,$\sigma$ for the rest. The best-fit occultation depths of the individual visits are $46^{+13}_{-14}$, $77^{+14}_{-13}$, $105^{+13}_{-14}$, $102^{+13}_{-14}$, $71^{+15}_{-16}$, $125^{+18}_{-19}$, and $51\pm15$ ppm. They are shown in Fig.~\ref{fig:variability} along with their posterior distributions in the side panel. The scatter of these measurements points towards a possible variability in the occultation depth of KELT-20b. The individual measurements are up to $2.8\,\sigma$ discrepant from their weighted mean, with the two most extreme measurements disagreeing by $3.4\,\sigma$. Conscious that the small sample size complicates a thorough statistical evaluation, we nevertheless attempted to quantify its significance. To do so, we created $10^4$ mock datasets assuming a constant occultation depth and measurements scattered around it following a Gaussian distribution with a width equal to the measurements' uncertainties. From this distribution, we find a 0.3\% probability (i.e.\ a p-value of 0.0033, or a $2.9\,\sigma$ significance) that measurements at least as discrepant as those observed are created randomly. While encouraging, this value hinges on the exact placement and uncertainty of each individual measurement, which hinges on the analysis approach (see Appendix \ref{app:depths}), and the treatment of stellar and instrumental correlated noise. The leftover correlated noise in the best-fit residuals contributes $\sim20\%$ to the uncertainty levels,  the rest being the white noise. Considering this non-negligible red noise to some extent could contribute to the observed discrepancies in the occultation depth. On the other hand, it is noteworthy that the simultaneous TESS data (averaged in each sector) show a similar behaviour at lower significance, tentatively supporting the variability. Using all the visits, we obtain a mean occultation depth of 82$\pm$6 ppm, as shown by its posterior in the right plot of Fig. \ref{fig:bestfit_transit_occultation}. We also present the individual occultation best-fits in the right panel of Fig. \ref{fig:bestfits_cheops}. 

After finding  the variability in the \cheops\ occultation depths, we looked for signs of variability in the TESS observations, which   simultaneously observed the target during its Sector 40 and 41 orbits. The phase curve analysis from the second year of TESS has already been studied by \citet{Wong2021}, and based on the four occultations in Sector 14, they report a depth of $111^{+35}_{-36}$ ppm.  \citet{Fu2022ApJ} has analysed the occultation cuts from the three sectors observed at the time, and reports an average occultation depth of $139\pm8$ ppm. Following the same approach, we eliminated the data points centred at the individual occultations that span over three times the duration. We then fitted the data with the occultation model along with a linear detrending term. Upon respective sector-wise analysis, we obtained  occultation depths of $123\pm36$, $105\pm23$,  $196\pm32$, and $169\pm29$ ppm. Consequently, we observe a 2~$\sigma$ difference in the measured depths between Sectors 40 and 41. This approach is convenient for obtaining an average measurement. However, a sector-wise comparison would need a complete planetary model with a better characterisation of the systematics present in the light curve. 

\subsection{\normalfont{\tess} phase curves} \label{sec: phase curves}
We re-analysed the TESS light curves and fitted them with the complete model described in Section \ref{sec: Planetary Model}. We first fitted the model without any red noise correction term, and as a result  we found strong short frequency trends in the residuals. To account for these low-frequency signals we simultaneously fitted the model with a Gaussian process (GP) model using a Mat\'ern 3/2 kernel \citep{Rasmussen2006, Gibson2012}. The Mat\'ern 3/2 kernel defines the covariance between two observations at times $t_{i_{1}}$ and $t_{i_{2}}$ as 

\begin{equation}
k(t_{i_1},t_{i_2})=h^2\left(1+\frac{\sqrt{3}|t_{i_1}-t_{i_2}|}{\lambda}\right)\exp\left(-\frac{\sqrt{3}|t_{i_1}-t_{i_2}|}{\lambda}\right)+j_o^2\delta_{i_1,i_2},\label{eq:kernelTESS}
\end{equation}
\noindent
where $h$ is the amplitude of the GP and $\lambda$ is its timescale. In order to take into account any additional white noise not included in the uncertainties, either instrumental or astrophysical,  we added the diagonal terms $j_o^2\delta_{i_1,i_2}$ to the Mat\'ern 3/2 kernel, where $o$ denotes the TESS Sectors 14, 40, 41, and 54.  Unfortunately, the precision in the \tess\ light curves is insufficient to provide precise constraints on the sector-wise phase offset and nightside flux. Phase offset and nightside flux convergence was reached only in the case of Sector 40 with values consistent with zero (i.e. $-11\pm32$ degrees; see Fig. \ref{fig: A_night_Ph_offset}). Prior constraints on these parameters from the analysis of the  Sector 14 phase curves were also consistent with zero \citep{Wong2021}. Hence, to reduce the dimensionality of the fit, we set these two parameters to zero. The parameters of the individual sector fits are presented in Table \ref{tab: TESS_sectorwise_phase_curve}. The occultation depths measured with this approach are in agreement with the value measured in the previous section, although these depths lie towards the shallow end of the 1\,$\sigma$ distribution obtained using the previous approach.

We fitted   all   four sectors, first independently and then simultaneously, with the same overall model parameters except the GP model hyper-parameters $h$ and $\lambda$,  which are specific to correlated trends in individual sectors, although we kept a separate jitter term to account for the white noise. All the model parameters described in section \ref{sec: Planetary Model} were kept free. The results of the best-fit parameters are presented in   Table \ref{tab:results}. The occultation depth is consistent with the previous analysis and literature values (Sect. \ref{sec: results_occ}). The best-fit model is shown in Figure \ref{fig:bestfit_transit_eclipse_TESS} and the sector-wise independent model plots are shown in Fig. \ref{fig: TESS_sector-wise}. We did not detect any significant nightside emission as  $A_{\rm night}$ is consistent with zero within 1~$\sigma$. We report a 99.9\% upper limit on $A_{\rm night}$ of 29 ppm, which corresponds to an upper limit to the nightside temperature at 2280\,K. We also did not find any significant phase offsets as $\Delta \phi$ is consistent with zero.  
\section{Discussion}\label{sec: discussion}
\subsection{Atmospheric modelling}\label{sec: Retrieval}
Given the bulk properties of {\keltb}, we expect that   thermal emission and stellar reflected signals both contribute to the planetary flux at the optical wavelengths probed by CHEOPS and TESS.

To provide a physical interpretation of the observed occultation depths, we modelled the planet's emission spectrum using the open-source {\pyratbay}
framework\footnote{\href{https://pyratbay.readthedocs.io/}{https://pyratbay.readthedocs.io/}}
\citep[][]{CubillosBlecic2021mnrasPyratBay}.
This package employs parametric models of the planetary atmosphere (1D pressure profiles of the temperature, composition, and altitude), from which we then generate the thermal emission spectra.  Using a Bayesian retrieval approach, the code constructs posterior distributions for the 
atmospheric parameters by fitting the emission models to observed occultation depth, using a differential-evolution MCMC \citep{CubillosEtal2017apjRednoise}.  Since the {\pyratbay} code does not account for reflected light, we opted to constrain the
thermal-emission models only by the infrared observations, which are largely dominated by the thermal component.  We then computed the
predicted thermal emission in the CHEOPS and TESS bands, and
considered the reflected component as the difference between the
observed and modelled fluxes.

We modelled the atmosphere from 100 to $10^{-9}$~bar, adopting a
 configuration similar to  that of \citet{Fu2022ApJ}.  For the
temperature profile we used the model of
\citet{Guillot2010aaRadiativeEquilibrium}, parametrised by the mean
thermal opacity \textbf{$\log_{10}(\kappa')$}, the ratio of  visible and thermal
opacities \textbf{$\log_{10}(\gamma)$} to the irradiation temperature $T_{\rm irr}$.
We assumed an atmospheric composition in chemical equilibrium,
including the main species that dominate the chemistry for the
molecules expected to be probed by the observations.  These include
{\molhyd}, He, {\water}, {\methane}, CO, {\carbdiox}, N$_2$, HCN,
{\ammonia}, {\acetylene}, C$_2$H$_4$, TiO, VO, TiO$_2$, VO$_2$, S$_2$,
SH, SiO, H$_2$S, SO, and SO$_2$, as well as atomic and ionic species.
To vary the atmospheric composition in the retrieval models, we used
as parameters the abundance of metals [M/H], carbon [C/H], and oxygen
[O/H], all  relative to solar abundances in logarithmic scale.
For these parameters we used a uniform prior ranging from 0.01 to 100
times solar (Table \ref{table:retrieval}).

\begin{figure*}[t]
\centering
\includegraphics[width=\linewidth]{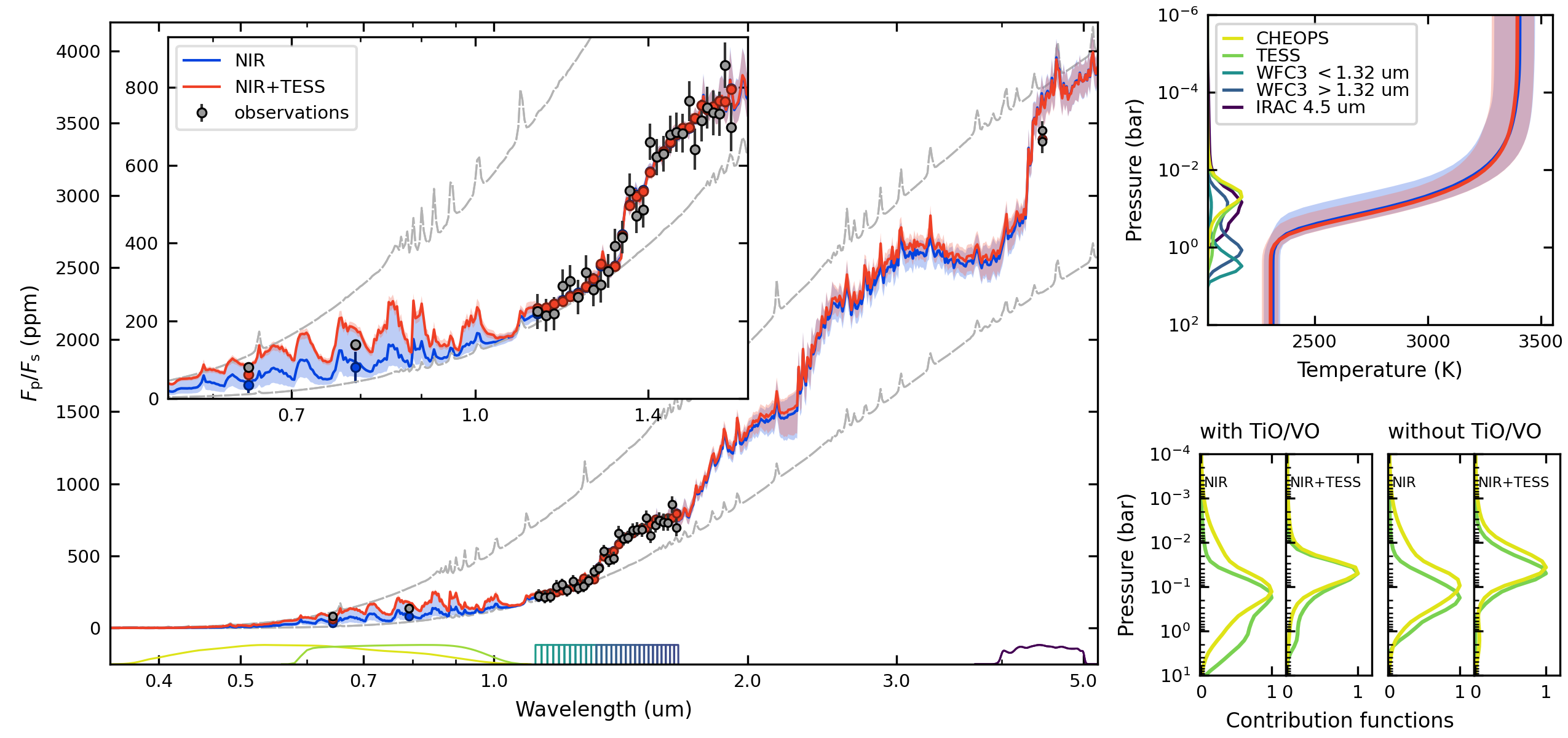}
\caption{\textbf{\textit{Left:}} {\keltb} observed occultation depths (grey
  markers). The curve at the bottom shows the through-puts for each data
  point.  The solid curves show the median of the retrieved model
  distributions when fitting the NIR {\HST} and {\Spitzer}
  occultations (blue) and the NIR+TESS occultations (red).  The
  shaded areas (in corresponding colours) denote the span of the 68\%
  central percentile of the distributions. The dashed grey curves show
  spectra for two black-body planetary model spectra at 2300 and
  3200~K. \textbf{\textit{Top right:}} Retrieved temperature profiles (same colour-coding as before). The curves at the left edge show the contribution
  functions for selected bands (see legend). \textbf{\textit{Bottom right:}}
  Contribution functions integrated over the CHEOPS and TESS
  bands for the two cases shown in the left panel and for retrievals without TiO and VO opacity (same colour-coding as above). The band-integrated thermal flux in {\cheops} and {\tess} bands are ($36\pm18$ ppm) and ($85\pm38$ ppm), respectively (coloured markers).}
\label{fig:retrieval_atmosphere}
\end{figure*}

We computed the emission spectra for each atmospheric model with a
line-by-line radiative transfer routine sampling between 0.3 and 5.5
${\microns}$ at a constant resolving power of
$R=\lambda/\Delta\lambda=15\,000$.  For the molecular opacities we used
the data from HITEMP for CO, {\carbdiox}, and {\methane}
\citep{RothmanEtal2010jqsrtHITEMP, LiEtal2015apjsCOlineList,
  HargreavesEtal2020apjsHitempCH4}, and from ExoMol for {\water},
{\ammonia}, HCN, TiO, and VO \citep{HarrisEtal2006mnrasHCNlineList,
  HarrisEtal2008mnrasExomolHCN,
  PolyanskyEtal2018mnrasPOKAZATELexomolH2O,
  ColesEtal2019mnrasNH3coyuteExomol,
  Yurchenko2015jqsrtBYTe15exomolNH3,
  McKemmishEtal2016mnrasVOMYTexomolVO,
  McKemmishEtal2019mnrasTOTOexomolTiO}. Prior to the retrieval we
pre-processed these large data files, first using the {\repack}
algorithm to extract the dominant line transitions
\citep{Cubillos2017apjRepack}, and then tabulating the  opacities 
over a fixed pressure, temperature, and wavelength grid.
In addition, the model included Na and K resonant-line opacities
\citep{BurrowsEtal2000apjBDspectra}; collision-induced opacities for
{\molhyd}--{\molhyd} \citep{BorysowEtal2001jqsrtH2H2highT,
  Borysow2002jqsrtH2H2lowT}; Rayleigh-scattering opacity for H,
{\molhyd}, and He \citep{Kurucz1970saorsAtlas}; and a grey cloud deck
model.

Figures \ref{fig:retrieval_atmosphere} and
\ref{fig:retrieval_posteriors} and Table \ref{table:retrieval}
summarise our atmospheric retrieval results. To account for the
unknown impact of reflected emission in the optical, we considered two
retrieval scenarios: one fitting only the {\HST}/WFC3 and
{\Spitzer}/IRAC near-infrared (NIR) occultation depths of
\citet{Fu2022ApJ}, and another that fits the NIR data and TESS
occultation depth reported in this work.

The retrieval models fit the occultation data well, with the {\HST} data dominating the thermal structure constraints.  The higher emission seen in the longer wavelengths
1.3--1.6~$\micron$, where {\water} absorbs more strongly, drives the retrieval towards inverted
temperature profiles.  In terms of composition, both fits (with and without the TESS constraint) find a strong correlation between the carbon and oxygen abundances (Fig.\ \ref{fig:retrieval_posteriors}). This results in broad carbon and oxygen posteriors, but a well-defined  C/O ratio of less than one:  C/O$ = 0.88^{+0.08}_{-0.23}$ (NIR retrieval)
and C/O$ = 0.72^{+0.19}_{-0.22}$ (NIR+TESS retrieval).
Including the TESS constraint has the largest impact on the [M/H] parameters, which accounts for the metallicity of all other species (most affected by the  TESS constraint) and which narrows down to solar values from an otherwise largely unconstrained posterior. This occurs because the TESS band probes the absorption from species
such as TiO, VO, Na, and K, which were previously unconstrained.  The relatively high value of the TESS occultation depth drives the [M/H] constraint towards the upper values given the non-inverted thermal profile (higher emission from the hotter, upper layers of the atmosphere). Our results are generally consistent with those of \citet{Fu2022ApJ};
 both analyses find a clear thermally inverted temperature profile that quickly increases from $\sim$2300\,K to $\sim$3400\,K and C/O ratios close to the solar value (C/O$_\odot=0.55$). In one additional test we explored cases excluding the absorption from TiO and VO, which are molecules that tend to condense out of planetary atmospheres
\citep[e.g.,][]{HoeijmakersEtal2020aaWASP121bHiresSpecies}.  For these runs we found no significant variations in the retrieved properties (results consistent at the $1\sigma$ level).

In terms of thermal-light {versus} reflected-light flux,  our NIR
retrieval underestimates the TESS and CHEOPS occultation depths
(Fig.\ \ref{fig:retrieval_atmosphere}), indicating that there must be a
non-negligible reflected-light component.  By including the TESS
measurement in the fit, the model matches the TESS occultation depth,
but still underestimates the CHEOPS occultation depth, again
suggesting that there should be some fraction of reflected light in
that band,  although we note that this fit makes the implicit
assumption that the TESS occultation depth is purely due to thermal
emission. Therefore, these two scenarios model the planetary
optical emission under two extreme cases: no prior assumption of
reflected or thermal contribution (NIR retrieval) {versus} maximum
thermal contribution (NIR+TESS retrieval).
Our contribution functions indicate that TESS and {\Spitzer}
probe higher layers in the atmospheres than {\HST}.  In each scenario,
the CHEOPS contribution functions coincide fairly well with that of
TESS, suggesting that they probe a similar region in the atmosphere
(Fig.\ \ref{fig:retrieval_atmosphere}).

{\renewcommand{\arraystretch}{1.3}
\begin{table}
\begin{minipage}{\linewidth}
\centering
\caption{{\keltb} Atmospheric retrieval. The reported values and
  uncertainties correspond to the median and 68\% quantile of the
  marginal posterior distributions, respectively.}
\label{table:retrieval}
\begin{tabular*}{\linewidth} {@{\extracolsep{\fill}} lccc}
\hline
Parameter           & Priors                 & Retrieved              & Retrieved  \\
                    &                        & without TESS              & with TESS \\
\hline
$\log_{10}(\kappa')$ & $\mathcal U(-8, -1)$   & $-5.97^{+0.51}_{-0.21}$ & $-6.02^{+0.37}_{-0.17}$ \\
$\log_{10}(\gamma)$ & $\mathcal U(-4, 2)$     & $0.67^{+0.04}_{-0.06}$  & $0.67^{+0.04}_{-0.06}$ \\
$T_{\rm irr}$ (K)   & $\mathcal U(1000, 3500)$ & $2618^{+35}_{-40}$     & $2610^{+35}_{-40}$ \\
$[$M/H$]$          & $\mathcal U(-2, 2)$      & $-0.79^{+0.75}_{-0.82}$  & $-0.08^{+0.13}_{-0.10}$ \\
$[$C/H$]$          & $\mathcal U(-2, 2)$      & $-0.58^{+1.09}_{-0.63}$ & $-0.74^{+0.94}_{-0.56}$ \\
$[$O/H$]$          & $\mathcal U(-2, 2)$      & $-0.73^{+1.03}_{-0.52}$ & $-0.82^{+0.84}_{-0.40}$ \\
\hline
\end{tabular*}
\end{minipage}
\end{table}
}

\subsection{Albedo}
\subsubsection{Geometric albedo and brightness temperature} \label{sec: Albedo}
\begin{figure*}[h!]
\centering
\includegraphics[width=.55\hsize,trim={0 0cm 0cm 0cm},clip]{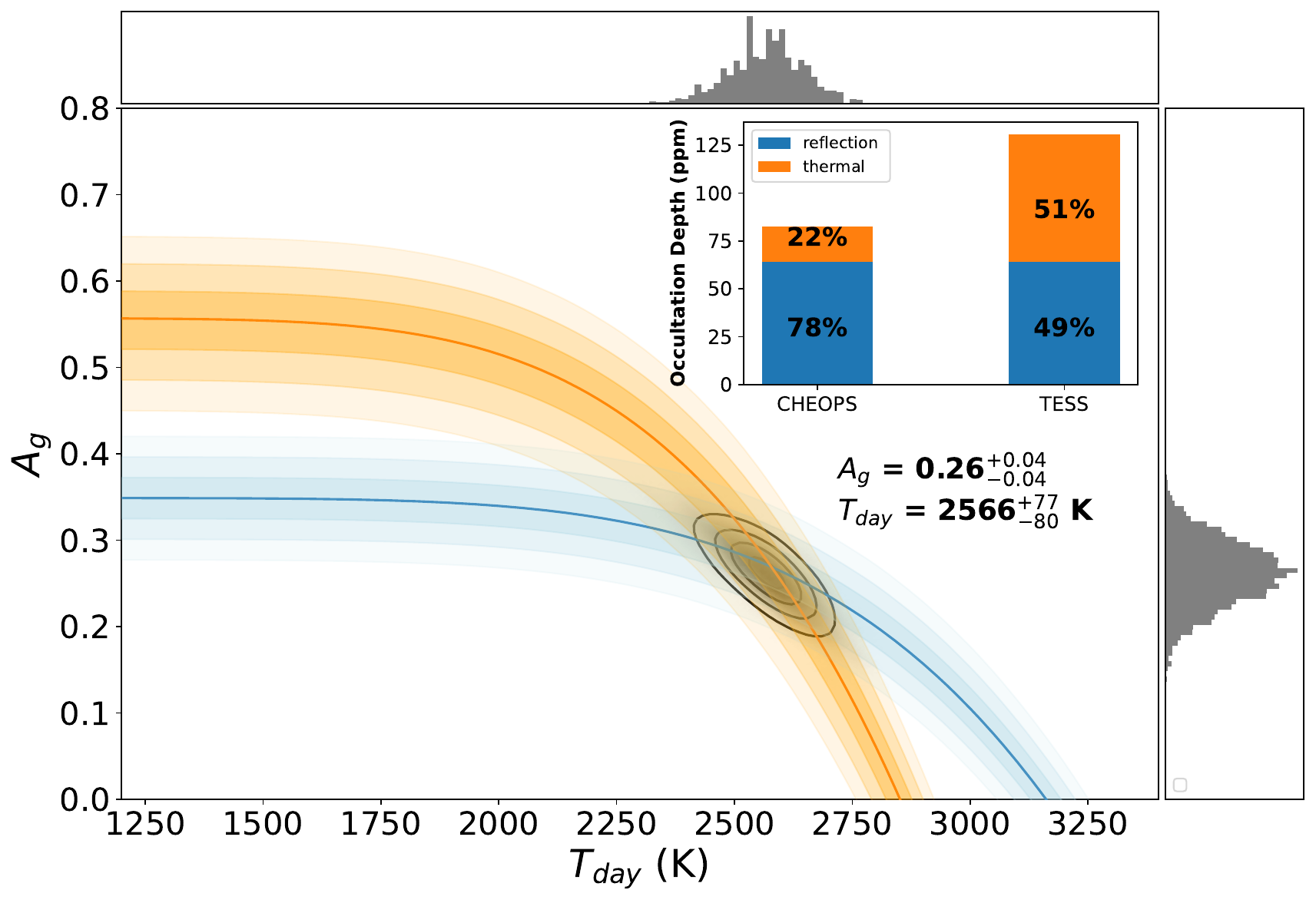}
\includegraphics[width=.39\hsize,trim={0cm 0cm 0.5cm 0.5cm},clip]{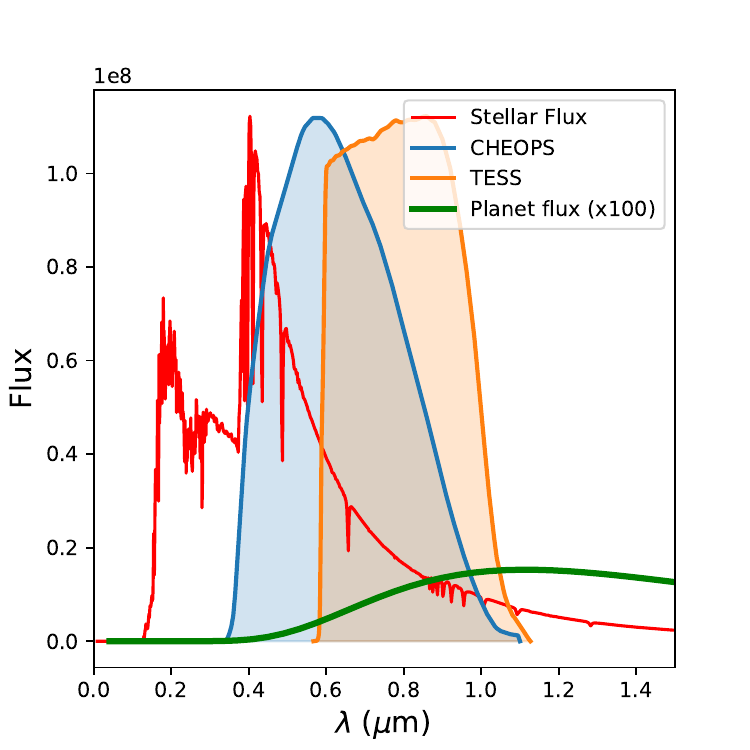}
\caption{Dayside reflection and emission brightness contributions. \textbf{\textit{Left:}} Geometric albedo $(A_{g})$ as a function of the dayside brightness temperature for the estimated occultation depths in CHEOPS (blue) and TESS (red). The decreasing intensity of the colour shades corresponds to the 1\,$\sigma$, 2\,$\sigma$, and 3\,$\sigma$ uncertainties in the computation of $A_{g}$ for each $T_{\mathrm{day}}$.  The overlapping region provides the true constraints on the two parameters represented by the posterior distribution in grey. The individual posteriors of $(A_{g})$ and $(T_{\mathrm{day}})$ are presented in the side panels. The resultant values are reported in the text. Subpanel: Reflection and emission contributions to the occultation depth of the CHEOPS and TESS passbands. The identical height of the reflection component reflects the assumption of $\alpha=1$. 
\textbf{\textit{Right:}} Kurucz stellar model [$T_{eff}=9000$, $[Fe/H]=0$ and $logg=4.5$] along with the {\cheops} and {\tess} passband responses. The dayside planetary black-body flux (x100) is also plotted for a comparison with the stellar flux, $T_{day}=2600$~K.}
\label{fig: Ag_vsTday}
\end{figure*}

A geometric albedo ($A_{g}(\lambda)$ or $A_{g,\lambda}$) is defined as the spherical albedo at full phase or phase angle $\alpha=0$ \citep[superior conjunction,][]{Seager2010exop.book}. It is defined as the ratio of the planet's brightness to that of an idealised, flat, fully reflecting Lambertian disc with an identical cross-section. The brightness temperature ($T_{\mathrm{day}}(\lambda)$) of the planetary day side is the temperature of a black-body that emits the same amount of radiation within a specific passband. The T-P profile from Fig. \ref{fig:retrieval_atmosphere} suggests that \cheops\ and \tess\ practically probe the same layer of the atmosphere that irradiates at similar temperatures, which are computed to be $2759^{+200}_{-250}$ and $2555^{+200}_{-250}$ K, respectively. Alternatively, following the {\HST}/WFC3 + {\Spitzer}/IRAC (NIR) model, we computed the band-integrated emission flux for both CHEOPS and TESS at 36$\pm$18 ppm and 85$\pm$38 ppm, respectively. As the resultant occultation depth is the sum of the two theoretical depths (i.e. $\delta_{\mathrm{occ}} = \delta_{\mathrm{ref}} + \delta_{\mathrm{thm}}$). Subtracting the emission flux from the observed depths listed in Table \ref{tab:results}, we obtain the reflected flux and using Eq. \ref{eqn:Albedo} \citep[]{Seager2010exop.book}, we obtain the band-integrated geometric albedos in the \cheops\ and \tess\ bands as $A_{g,{\mathrm{CHEOPS}}} = 0.19 \pm 0.08$ and $A_{g,{\mathrm{TESS}}} = 0.19 \pm 0.16$, respectively: 
\begin{equation}\label{eqn:Albedo}
\delta_{\rm ref} = A_{g}\left(\frac{R_{p}}{a}\right)^{2}.
\end{equation}
The geometric albedos in the two passbands are consistent with each other, as are the brightness temperatures from the T-P profile. Moreover, the retrieval model does not include scattering. In the absence of a detailed occultation spectrum at optical wavelengths, it is difficult to precisely predict the exact contribution of thermal emission or reflection to the planetary brightness. Therefore, to solve this degeneracy problem and to obtain stricter constraints on the dayside brightness temperature and the geometric albedo, we explored an alternate approach. Following the definition, we express the dayside emission in terms of the  brightness temperature as Eq. \ref{eqn:Emission}  \citep[]{Santerne2011A&A, Singh2022}:

\begin{equation}\label{eqn:Emission}
    \delta_{\rm thm} =
    \pi\left(\frac{R_{\rm p}}{R_{\star}}\right)^{2}  \frac{ \int_{\lambda} B_{\lambda}(T_{\mathrm{day}})   \Omega_{\lambda} d\lambda}{\int_{\lambda} S_{\lambda}^{\rm CK} \Omega_{\lambda} d\lambda}\,.
\end{equation}
\noindent
Here $B_{\lambda}(T_{\mathrm{day}})$ is the Planck distribution for temperature $T_{day}$, $S_{\lambda}^{\rm CK}$ is the stellar Kurucz flux \citep{Castelli&Kurucz2003IAUS} for the nearest $T_{\rm eff}$, $\log{g}$, and [Fe/H]. The stellar flux and the planetary black-body flux is integrated over the telescope's passband $\lambda$ with the corresponding response function $\Omega_{\lambda}$\footnote{Telescope response functions were obtained from the SVO filter profile service:  http://svo2.cab.inta-csic.es/theory/fps/}. We use the measured value of $\delta_{\mathrm{occ}}$ from observations together with Equations \ref{eqn:Albedo} and \ref{eqn:Emission} to create a parametric relationship between the geometric albedo and the brightness temperature as 

\begin{equation}\label{eqn: Albedos}
    \frac{\alpha\delta_{\mathrm{occ}}^{\mathrm{CHEOPS}} - \delta_{\mathrm{occ}}^{\mathrm{TESS}}}{\left(\frac{R_{p}}{a}\right)^2} + \delta_{\mathrm{thm}}^{\mathrm{TESS}} - \alpha\delta_{\mathrm{thm}}^{\mathrm{CHEOPS}} = 0,
\end{equation}
\noindent
where $\alpha$ denotes the ratio of the two geometric albedos (i.e. $A_{g}^{\mathrm{TESS}}/A_{g}^{\mathrm{CHEOPS}}$). Assuming identical band-integrated geometric albedos (or $\alpha=1$) and the same brightness temperatures in the two passbands, we can precisely constrain the two $A_{g}-T_{\mathrm{day}}$ pairs. Figure~\ref{fig: Ag_vsTday} shows the $A_{g}$ versus $T_{\mathrm{day}}$ relation for the occultation depths measured in the \cheops\ and \tess\ passbands. The brightness temperature is the independent parameter that varies from 1200\,K to 3500\,K. We do not expect negative values for the geometric albedo and are therefore truncated in the plot.  

Being a wavelength-dependent property, $A_{g}$ might, in principle, vary between the \tess\ and \cheops\ bands (i.e. $\alpha \neq 1$); however, to constrain the $T_{\mathrm{day}} - A_{g}$ pair, we assume this variation to be below the precision of our data. This corresponds to one of  two scenarios: (1) a grey-sky reflective atmosphere (i.e. a flat reflection spectrum) or (2) if the albedo spectrum happens to be non-monotonic and gets averaged out to produce an identical band-integrated $A_{g}$ for the two passbands. Following this methodology we obtained precisely constrained values of $A_{g}$ and $T_{\mathrm{day}}$ at $0.26\pm0.04$ and $2566_{+77}^{-80}$\,K, respectively. We obtain a  dayside temperature similar to that measured in Section \ref{sec: Retrieval}. Furthermore, we derive the respective contributions of reflection ($\delta_{\mathrm{ref}}$) and emission ($\delta_{\mathrm{thm}}$) to the occultation depth and are shown in the subfigure in Fig.~\ref{fig: Ag_vsTday}. In the {\cheops} band, the reflection--emission contribution is almost 80$\%$--20$\%$, while in the {\tess} band, they are equally distributed. The derived brightness temperature and albedo posteriors along with the posteriors of the fitted parameters used in their computation are shown in Fig.~\ref{fig: Ag_Tday_posterior}. $A_{g}$ and is more precisely constrained compared to the previous approach. $A_{g}$ correlates with the occultation depth in \cheops\ as visible in the posterior distribution. The agreement between the $A_{g}$ values determined with the two approaches is within 1\,$\sigma$. On the other hand the brightness temperature correlates with the occultation depth measured in \tess.

Additionally, the same dataset can also be explained with alternative scenarios. If we break the identical $A_{g}$ assumption, but keep the same $T_{\mathrm{day}}$, then the maximum brightness temperature is $2857\substack{+29\\-25}$ K, which is  equal to the maximum $T_{\mathrm{day}}$ associated with $\delta_{\mathrm{occ}}$ of \tess. This would imply $A_{g,\mathrm{TESS}} = 0$ and $A_{g,\mathrm{CHEOPS}} = 0.17\pm0.03$ (i.e. entirely absorbing atmosphere in \tess\ band, and yet significantly reflective in the \cheops\ band). Furthermore, if we assume no reflection contribution whatsoever, then we need to give up the black-body assumption. This leads to maximum $T_{\mathrm{day}}$ associated with the observed $\delta_{\mathrm{occ}}$ of {\cheops}, $T_{\mathrm{day}}$ = $3162\substack{+45\\-47}$\,K. Nevertheless, this scenario is less likely following Sect.~\ref{sec: Retrieval}. 

\begin{figure}[h!]
\centering
\includegraphics[width=\linewidth]{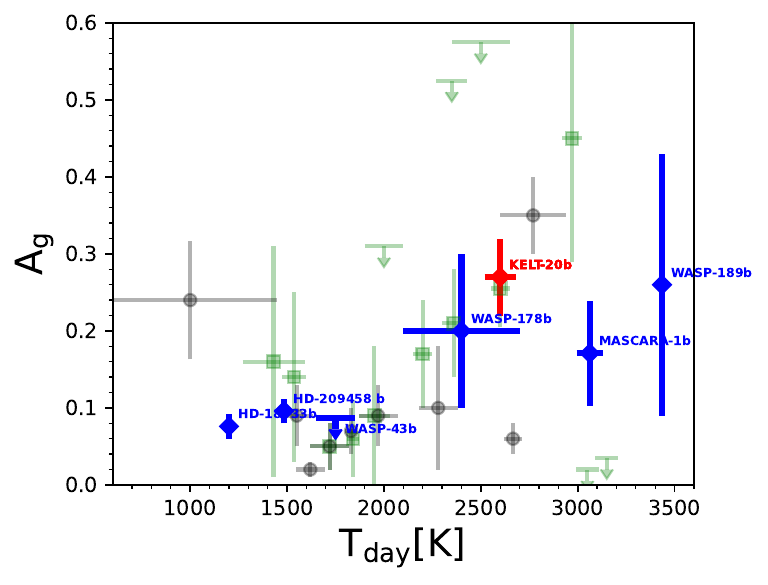}
\hspace{1.0cm}
\caption{$A_{g} - T_{day}$ of several hot Jupiters following \cite[]{Wong2021}. The black circles (\textit{Kepler}-\textit{CoRoT})  and the green squares (\tess)  are taken directly from their sample. The blue diamonds correspond to the published values of the \cheops\ hot Jupiters \citep{Krenn2023A&A, Brandekar2022A&A, Scandariato2022A&A, Pagano2023arXiv, Hooton2022A&A, Deline2022A&A}. The red diamond is from this work on \keltb. }
\label{fig:WongPlot}
\end{figure}

Following \cite[]{Wong2021}, we re-plotted the $A_{g}$ versus $T_{\mathrm{day}}$ for the sample of the hot Jupiters discussed in their paper along with the published \cheops\ results in Fig.~\ref{fig:WongPlot}. It is quite clear that there is  a positive correlation between $A_{g}$ and $T_{\mathrm{day}}$ for HJ to UHJ in their work, with the exception of a few outliers like WASP-18b and WASP-33b. We observed a similar trend within the \cheops\ sample, thereby corroborating   that UHJs tend to be more reflective compared to HJs. Although in comparison with the entire sample, the addition of the {\cheops} sample suggests a flattening of A$_{g}$ for higher temperatures. 

\subsubsection{Bond albedo}
The Bond albedo \citep[$A_{B}$,][]{Bond_1861MNRAS} or the spherical bolometric albedo \cite[]{Mallama2017} is a measurement of the fraction of the total power of the incident radiation that is scattered back at all wavelengths and all phase angles. It provides a measure of a planet's energy budget that further provides information about its atmospheric composition and the overall thermal properties \citep[]{Schwartz2015MNRAS}. To convert a geometric albedo into a Bond albedo, we need information about the planet's reflectance spectrum, scattering phase function, and spatial inhomogeneity \citep[]{Hanel2003essi.book.....H}. In practice, the geometric albedo is related to the spherical albedo through the phase integral \textit{q} (i.e. the integral of the scattering phase function), which  gets translated to the Bond albedo through the  relation

\begin{equation}\label{eqn: Spherical Bond albedo}
     A_{\rm B}=\frac{\int_0^\infty q(\lambda)A_{\rm g}(\lambda)I_\star(\lambda)d\lambda}{\int_0^\infty I_\star(\lambda)d\lambda},
\end{equation}
where  $I_{\star}$ is the wavelength-dependent stellar intensity. For a Lambert sphere, q=2/3, and assuming a geometric albedo independent of wavelengths (i.e. a flat albedo spectrum), we obtain $A_{B} = 3/2A_{g}$. This formalism has been the norm during the \textit{Kepler} and the {\Spitzer} era. 

Following \cite[]{cowan2011statistics}, the effective dayside ($T_{d}$)\footnote{\cite{Seager2010exop.book} uses the same formulation for the equilibrium temperature of the day side. The effective temperature is the equilibrium temperature if the heat flux from the planet's interior is negligible. They are used interchangeably in the literature.} and nightside temperatures ($T_{n}$) can be expressed in terms of the bond albedo $A_{B}$ and the heat re-circulation efficiency $\epsilon$ that parametrises the heat flow from the day side to the night side (Eq. \ref{eqn:Dayside} and \ref{eqn:Nightside}): 
\begin{equation}\label{eqn:Dayside}
T_{d} = T_{\rm eff} \sqrt{\frac{R_{\star}}{a}} \left(1 - A_{B}\right)^{\frac{1}{4}} \left(\frac{2}{3} - \frac{5}{12}\epsilon \right)^\frac{1}{4}
,\end{equation}
\begin{equation}\label{eqn:Nightside}
T_{n} = T_{\rm eff} \sqrt{\frac{R_{\star}}{a}} \left(1 - A_{B}\right)^{\frac{1}{4}} \left(\frac{\epsilon}{4} \right)^\frac{1}{4}.
\end{equation}

\begin{figure*}[h!]
\begin{minipage}[c]{0.75\textwidth}
\begin{tikzpicture}
\centering
\node(a){\includegraphics[width=0.8\linewidth]{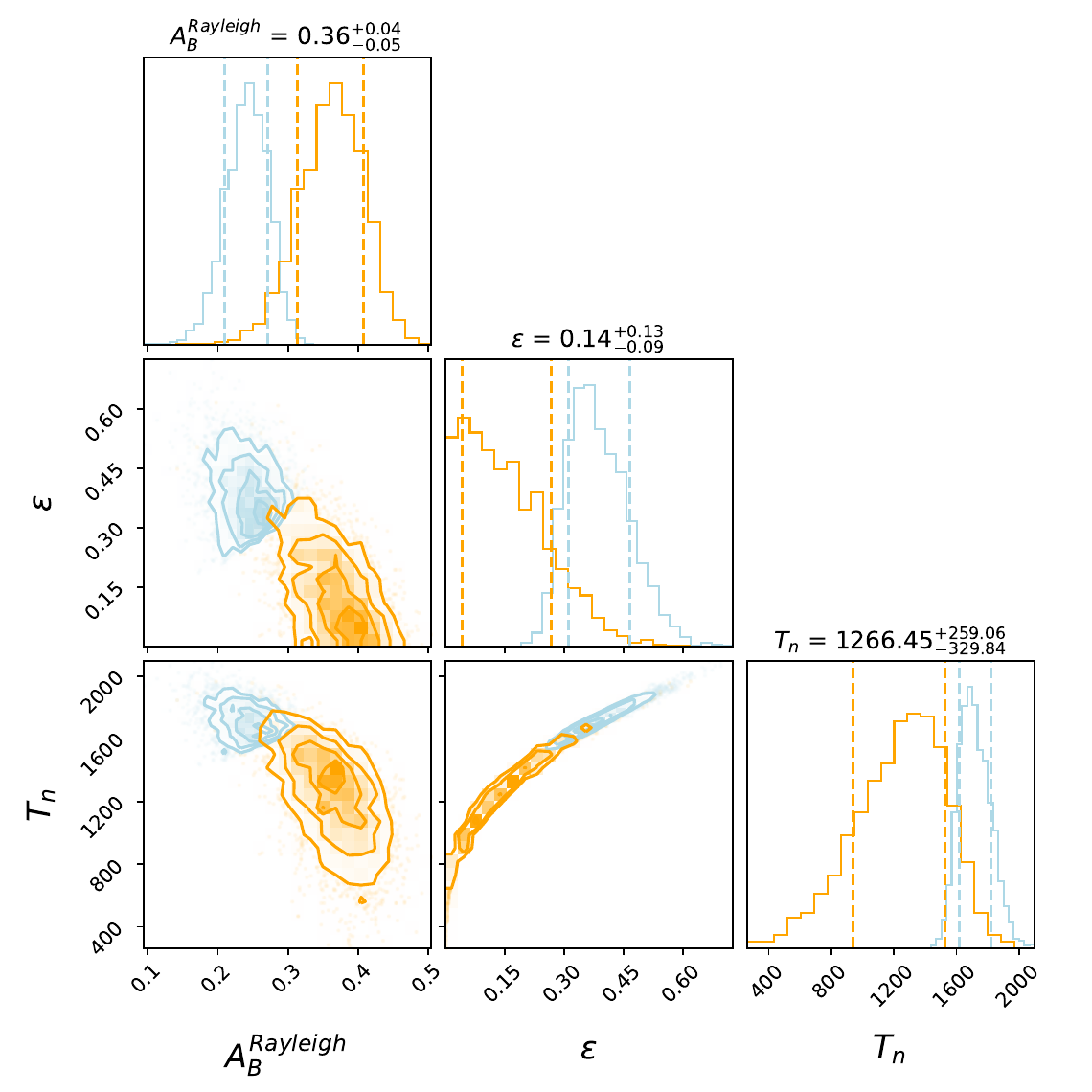}};
    \node at (a.north east)
    [
    anchor=north east,
    xshift=40mm,
    yshift=0mm
    ]
    {
    \includegraphics[width=.47\linewidth]{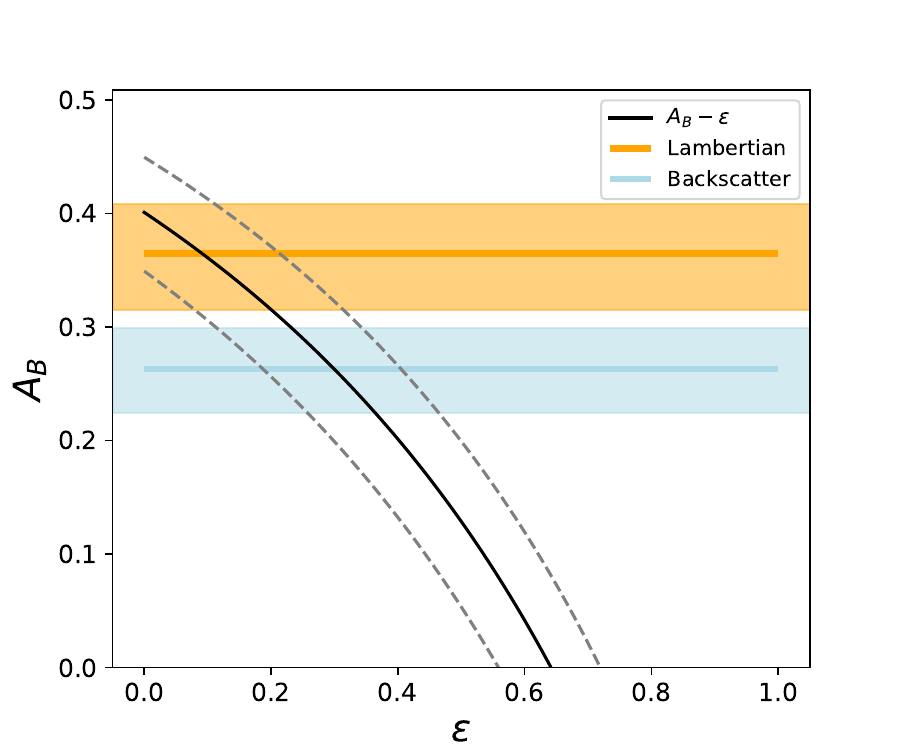}
    };
\end{tikzpicture}
\end{minipage}\hfill
\begin{minipage}[t]{0.25\textwidth}
\caption{Parameters determining planetary energy distribution under radiative equilibrium. \textbf{\textit{Top right:}} Bond albedo plotted as a function of the heat re-circulation efficiency for the effective dayside temperature $T_{\rm d}$. \textbf{\textit{Bottom left:}} Respective posteriors of $A_{B}$ and $\epsilon$ from the overlapping region, which are then used to compute the nightside temperature. The values in the label correspond to the orange posteriors, while the light blue posteriors are shown only for comparison.
}
\label{fig: Ab_epsilon}  
\end{minipage}
\end{figure*}

\noindent
Here $\epsilon = 0$ indicates instantaneous re-radiation or no re-circulation of heat from the day side to the night side, while $\epsilon = 1$ means perfect heat re-circulation. In the previous section we estimated the observed brightness temperature of the planet $T_{\mathrm{day}}$ within the CHEOPS-TESS passband. The best-fit planetary spectral energy distribution (SED) obtained from Sect. \ref{sec: Retrieval} equates to a thermal flux with an irradiation temperature of $2618^{+35}_{-40}$~K. We use this as the effective temperature of the planetary day side $T_{d}$. We therefore invert Eq. \ref{eqn:Dayside} to provide constraints on the associated bond albedo $A_{B}$ with varying $\epsilon$, as shown in Figure \ref{fig: Ab_epsilon}. The solid black line represents the $A_{B}$ value with varying $\epsilon$, while the dashed grey lines are the 16${th}$ and 84${th}$ percentiles. The orange and light blue span represents the constraints on the bond albedo following the  Lambertian scattering law (i.e. $q = 3/2$) and the  100$\%$ back-scattering law (i.e. $q = 1$), respectively. These estimates are computed using the $A_{g}$ value derived in   Sect. \ref{sec: Albedo}.

For the extreme case of inefficient heat re-circulation (i.e. $\epsilon=0$), we derive an upper limit on the bond albedo of $0.45\pm0.07$. We note that this does not make any assumptions regarding the phase integral ($q$), which relates the geometric albedo to the Bond albedo. We can however use our derived value of $A_{g}$ within the {\cheops}-{\tess} passbands to further constrain the limits on $A_{B}$ and $\epsilon$. Following the same approach as \cite[]{Schwartz2015MNRAS}, we use Eq. \ref{eqn: Spherical Bond albedo} and consider the two limits on the phase integral (i.e. q=3/2, Lambert scattering, and q=1, 100$\%$ back-scattering). As we have no information of the geometric albedo in the UV/FUV wavelength ranges, we assume a constant $A_{g}$ equal to that measured in the optical. Figure \ref{fig: Ab_epsilon} shows the posterior distribution of the $A_{B}$, $\epsilon$ along with the dayside and nightside temperatures for the two limiting scenarios.

We report the values corresponding to the flat $A_{g}$ with the planet following Lambertian scattering, which gives $A_{B} = 0.36^{+0.04}_{-0.06}$, $\epsilon$ = $0.14^{+0.13}_{-0.10}$ and the nightside temperature $T_{n} = 1266^{+259}_{-330}$~K. The uncertainties are propagated from the error estimates on the stellar effective temperature from Table \ref{tab: stellar parameters} and the semi-major axis from Table \ref{tab:results}. The low value of $\epsilon$ indicates weak heat re-circulation, while the average nightside temperature will result in emission signatures that too low  to be detectable in the \tess\ phase curves, which corroborates well with the nightside emission estimates that is consistent with zero. If a more asymmetric scattering law is assumed ($q<3/2$), a combination of smaller $A_{B}$ and higher $\epsilon$ would also be possible. In the limiting case of $q=1$ or $A_{B} = A_{g}$, we obtain $\epsilon$ as $0.38^{+0.09}_{-0.07}$, while the nightside temperature $T_{n}$ is $1708^{+113}_{-93}$~K. 

\subsection{Atmospheric variability}\label{sec: Variability}
In addition to a precisely constrained geometric albedo of the day side of \keltb\,, our data also shows a tentative detection of temporal variability in the occultation depth (see Sect. \ref{sec: results_occ}), which most plausibly can be attributed to the varying flux from the planetary day side. Prior reports of variability in a hot-Jupiter atmosphere were announced for HAT-P-7\,b \citep{Armstrong2016NatAs} and WASP-12\,b \cite{Wong_WASP12_2022AJ}. For HAT-P-7\,b, variability was observed in the phase curve shape, interpreted as changes in the location of the brightest point. Although this interpretation was later challenged and attributed to stellar super-granulation \citep{Lally2022AJ}. The behaviour of \kelt\ appears to differ, showing changes in the occultation depth itself which can be best illustrated when we analyse its variability within the first four visits of CHEOPS in conjunction with the simultaneous TESS Sector 40 and 41 occultations.

If we only consider the \tess\ Sector 40 observations combined with the first two simultaneous occultation observations of \cheops, we obtained the respective depths of 102$\pm$22 and 62$\pm$9~ppm for \tess\ and \cheops. Following the method described in Section \ref{sec: Albedo}, these values yield a grey-sky geometric albedo of 0.16$\pm$0.1. Conversely, if we consider only TESS Sector 41 and the later two occultations of CHEOPS (i.e. visits 3 and 4), then the respective depths are 149$\pm$34 and 103$\pm$10~ppm. This scenario yields a higher grey-sky geometric albedo of 0.31$\pm$0.1. Driven by the less discrepant \tess\ measurements, the derived average dayside temperatures remain compatible within 1\,$\sigma$. This suggests that a brightness variability through reflectivity, as expected from varying cloud cover, rather than temperature, is at play.

Understanding the cloud coverage and variability requires understanding the underlying 3D cloud coverage. Recently, \citet{Helling2023} provided an extensive grid of 3D cloudy models for warm to ultra-hot Jupiters. \keltb\ is an ultra-hot Jupiter that corresponds best to the 2200\,K hot 3D case around an F-type star in their Figure~5. Such a planet is expected to have a partly cloudy morning terminator because the relatively slow rotation of the tidally locked planet of 3.5~days still promotes the formation of a strong super-rotating jet, which carries cloud particles from the night side towards the day side, that can cover up to 30~degrees of the morning terminator with thick clouds in the steady-state climate. Therefore, the cloud coverage on the dayside limbs on such ultra-hot Jupiters is mostly determined by the strength of the super-rotating wind jet.
\par
The location of the hot spot shift is also highly sensitive to the strength of the super-rotating wind jet. We note, however, that compared to hot Jupiters like HD~209458\,b with an effective dayside temperature $T_{\mathrm{d}}=1500$\,K, the hot spot shift is very much reduced for an ultra-hot Jupiter with $T_{\mathrm{d}}>2000$\,K because the radiative timescales are proportional to $\propto T^{-3}$, and thus become very short compared to the dynamical timescale. In other words, the hot air on the day side cools off faster as it is transported eastwards towards the night side.

It   has been pointed out again by \citet{Helling2023} that such a planet is also highly thermally ionised at the day side (their Figure~14). Several authors \citep{Komachek2020ApJ,Beltz2022,Hindle2021} suggest that the coupling of ionised flow with the planetary magnetic fields can then lead to drastic changes in the 3D wind structure, sometimes entirely suppressing the super-rotating wind jet. A suppressed super-rotating wind jet results in a reduction of the cloud influx towards the dayside hemisphere and a disappearance of any hot-spot offset on such a highly irradiated planet.

Additionally, changes in the interior heat flux can also lead to fluctuations in the strength of the super-rotating wind jet and thus the morning cloud and hot spot shift \citep{Carone2020,Schneider2022,Komacek2022}. Thus, a combination of magnetic field interactions and possible interior effects can lead to variations in  the occultation depth, as the hot spot is shifted more or less away from the substellar point, and in the planet's albedo, due to the accompanying morning cloud variability. 

Next to the low S/N, distinguishing between reflection and emission in the \tess\ band is difficult as it is impractical to incorporate two degenerate phase amplitude terms to account for reflection and emission separately. Therefore, a single phase offset term would have to account for both the individual phase variations. A phase offset consistent with zero can still involve super-rotating jets accompanied by morning cloud because  of the complementarity of the reflection bright spot and the emissive hot spot in the phase curve. Compared to \tess, the  \cheops\ occultations are more sensitive towards the reflected stellar light that accounts for around three-fourths of the total planetary optical dayside flux (Fig.~\ref{fig: Ag_vsTday} inset). While CHEOPS data cover occultations only and are thus not well suited to deriving phase curve peak offsets, the object might possess a phase offset in the optical due to enhanced reflection from clouds near the dayside terminator. Future high-precision phase curves of \keltb\ at bluer wavelengths can shed more light onto this aspect. 

Unlike HAT-P-7\,b \citep{Armstrong2016NatAs}, the variability in the occultation depth cannot be attributed to super-granulation in the host-star \citep{Lally2022AJ}. Because \kelt\  is an A dwarf star, it  does not possess a convective layer, and as a result it is less likely that it exhibits granulation signatures. Thus, \keltb\ is ideally amenable to study 3D weather and variability due to interactions between a highly ionised day side and the planetary magnetic field.

\subsection{Spin-orbit alignment}\label{sec: spinorbit}

Our gravity-darkening analysis (Section~\ref{sec:grav_dark}, Table~\ref{tab:results}) confirms the previous Doppler tomography measurements (Table~\ref{tab:obliquity}) that \keltb\ orbits in a plane that is misaligned from the stellar equatorial plane by at most a few degrees. As \cite{Barnes2011} point out, there is a four-way degeneracy in the value of $\lambda$ obtained from gravity-darkened transits; we must rely on the Doppler tomography results to exclude the retrograde solutions.

We see no evidence for nodal precession in \kelt\, unlike in the Kepler-13 \citep{Szabo_2012, Szabo_2020} and WASP-33 \citep{Johnson_WASP33, Watanabe_WASP33, Stephan_WASP33} systems, for instance, where the nodal precession is fast enough to cause measurable changes in the impact parameter of the transits over a relatively short baseline. This is in line with theoretical expectations \citep{Iorio2011} given that the \kelt\ system has a near-zero obliquity.

\begin{table}
\caption{Comparison of the sky-projected obliquity values.}             
\label{tab:obliquity}      
\centering          
\begin{tabular}{r l}      
\hline\hline       
$\lambda$ /degrees & Source \\
\hline                  
$3.4\pm2.1$ & \cite{Lund2017AJ} \\
$0.6\pm4.0$   & \cite{Talens2018} \\
$1.6\pm3.1$ & \cite{Hoeijmakers2020} \\
$3.9\pm1.1$& CHEOPS$^*$ \\
$9\pm1$& TESS$^*$ \\
\end{tabular}\\
$^*$ Only the prograde solutions for $\lambda$ are listed here, although a retrograde orbit cannot be excluded by the light curves alone. 
\end{table}

\section{Conclusion}\label{sec: conclusions}
In this work, we present a detailed optical picture of the \kelt\ system. Using \cheops\ observations of five transits and seven occultations of \keltb, as well as four Sectors worth of \tess\ data, we obtained a precise planetary radius along with a more than  $3~\sigma$ detection of the planet's obliquity and  constraints on the star--planet alignment. The \cheops\ photometry found an aligned orbit for \keltb, which is in contrast to previous \cheops\ studies that have found strongly inclined orbits for planets orbiting other A-type stars \cite{Lendl2020, Hooton2022A&A, Deline2022A&A}. At the same time, we also probed the atmospheric properties of this UHJ. 

We used the emission spectrum obtained with \HST\ and \Spitzer\ from the  literature to provide a picture of the UHJ dayside thermal profile. We confirmed the observation of temperature inversion in the atmosphere of \keltb. Additionally, with the help of \tess\ photometry in synergy with \cheops\, we provide strong constraints on the geometric albedo at $0.26\pm0.04$ and the dayside brightness temperature at $2566^{+77}_{-80}$\,K. The high geometric albedo corroborates a known trend of strongly irradiated planets being more reflective. Such a combination of the  $A_{g}-T_{day}$ pair results in a reflection emission contribution of 3:1 and 1:1 to the occultation depth for \cheops\ and \tess,\ respectively. This further leads to informative constraints on the planetary bond albedo and the heat re-circulation efficiency. The day side is too hot to form clouds, and  therefore the precise nature of the high geometric albedo could either be attributed to strong scattering by atoms or ions on the day side or by reflective clouds around the planet's terminator. We explored the possible scenarios of cloud reflections.

The high precision of the \cheops\ occultations also reveals signs of variability in the occultation depth, suggesting variable brightness of the planetary day side. As \cheops\ measurements are dominated by reflected light, the temporal variability in occultation depth could be attributed to variations in the extent of advected nightside clouds onto the planetary limbs (see e.g.\ \citealt[]{Helling2021A&A,Adams2022ApJ} for a discussion on the impact of clouds at the limbs of hot Jupiters and how these may affect dayside albedos). These authors note, however, that their simple model cannot address the variations seen to date. Thus, clearly more work is needed to investigate the precise origin of the variations that our \cheops\ measurements revealed. A combination of magnetic field interactions and possible interior feedback can also lead to such a variation. Spectroscopic occultation or phase curve observations of \keltb\ especially in the UV/FUV will shed more light on the magnitude and nature of reflective sources in this UHJ. While spectroscopic observations in the infrared can shed more light on the energy budget and the physical processes governing this planet, making it ideally suited for follow-up observations with \textit{HST}, \textit{JWST}, and eventually \textit{ARIEL}.


\begin{acknowledgements}
CHEOPS is an ESA mission in partnership with Switzerland with important contributions to the payload and the ground segment from Austria, Belgium, France, Germany, Hungary, Italy, Portugal, Spain, Sweden, and the United Kingdom. The CHEOPS Consortium would like to gratefully acknowledge the support received by all the agencies, offices, universities, and industries involved. Their flexibility and willingness to explore new approaches were essential to the success of this mission. CHEOPS data analysed in this article will be made available in the CHEOPS mission archive (\url{https://cheops.unige.ch/archive_browser/}). 
VSi, GSc, GBr, IPa, LBo, VNa, GPi, RRa, and TZi acknowledge support from CHEOPS ASI-INAF agreement n. 2019-29-HH.0. The author acknowledges the support of Prof. Kevin Heng for his comprehensive discussions on atmospheres of exoplanets during the early days of CHEOPS.
P.E.C. is funded by the Austrian Science Fund (FWF) Erwin Schroedinger Fellowship, program J4595-N. 
ML acknowledges support of the Swiss National Science Foundation under grant number PCEFP2\_194576. 
This work was also partially supported by a grant from the Simons Foundation (PI Queloz, grant number 327127). 
S.G.S. acknowledge support from FCT through FCT contract nr. CEECIND/00826/2018 and POPH/FSE (EC). 
ABr was supported by the SNSA. 
LCa and CHe acknowledge support from the European Union H2020-MSCA-ITN-2019 under Grant Agreement no. 860470
(CHAMELEON). 
TWi acknowledges support from the UKSA and the University of Warwick. 
This work was supported by FCT - Funda\c{c}\~{a}o para a Ci\^{e}ncia e a Tecnologia through national funds and by FEDER through COMPETE2020 through the research grants UIDB/04434/2020, UIDP/04434/2020, 2022.06962.PTDC. 
O.D.S.D. is supported in the form of work contract (DL 57/2016/CP1364/CT0004) funded by national funds through FCT. 
NCSa acknowledges funding by the European Union (ERC, FIERCE, 101052347). Views and opinions expressed are however those of the author(s) only and do not necessarily reflect those of the European Union or the European Research Council. Neither the European Union nor the granting authority can be held responsible for them. 
YAl acknowledges support from the Swiss National Science Foundation (SNSF) under grant 200020\_192038. 
RAl, DBa, EPa, and IRi acknowledge financial support from the Agencia Estatal de Investigación of the Ministerio de Ciencia e Innovación MCIN/AEI/10.13039/501100011033 and the ERDF “A way of making Europe” through projects PID2019-107061GB-C61, PID2019-107061GB-C66, PID2021-125627OB-C31, and PID2021-125627OB-C32, from the Centre of Excellence “Severo Ochoa'' award to the Instituto de Astrofísica de Canarias (CEX2019-000920-S), from the Centre of Excellence “María de Maeztu” award to the Institut de Ciències de l’Espai (CEX2020-001058-M), and from the Generalitat de Catalunya/CERCA programme. 
S.C.C.B. acknowledges support from FCT through FCT contracts nr. IF/01312/2014/CP1215/CT0004. 
XB, SC, DG, MF and JL acknowledge their role as ESA-appointed CHEOPS science team members. 
CBr and ASi acknowledge support from the Swiss Space Office through the ESA PRODEX program. 
ACC acknowledges support from STFC consolidated grant numbers ST/R000824/1 and ST/V000861/1, and UKSA grant number ST/R003203/1. 
This project was supported by the CNES. 
The Belgian participation to CHEOPS has been supported by the Belgian Federal Science Policy Office (BELSPO) in the framework of the PRODEX Program, and by the University of Liège through an ARC grant for Concerted Research Actions financed by the Wallonia-Brussels Federation. 
L.D. is an F.R.S.-FNRS Postdoctoral Researcher. 
B.-O. D. acknowledges support from the Swiss State Secretariat for Education, Research and Innovation (SERI) under contract number MB22.00046. 
This project has received funding from the European Research Council (ERC) under the European Union’s Horizon 2020 research and innovation programme (project {\sc Four Aces}. 
grant agreement No 724427). It has also been carried out in the frame of the National Centre for Competence in Research PlanetS supported by the Swiss National Science Foundation (SNSF). DE acknowledges financial support from the Swiss National Science Foundation for project 200021\_200726. 
MF and CMP gratefully acknowledge the support of the Swedish National Space Agency (DNR 65/19, 174/18). 
DG gratefully acknowledges financial support from the CRT foundation under Grant No. 2018.2323 ``Gaseousor rocky? Unveiling the nature of small worlds''. 
M.G. is an F.R.S.-FNRS Senior Research Associate. 
MNG is the ESA CHEOPS Project Scientist and Mission Representative, and as such also responsible for the Guest Observers (GO) Programme. MNG does not relay proprietary information between the GO and Guaranteed Time Observation (GTO) Programmes, and does not decide on the definition and target selection of the GTO Programme. 
SH gratefully acknowledges CNES funding through the grant 837319. 
KGI is the ESA CHEOPS Project Scientist and is responsible for the ESA CHEOPS Guest Observers Programme. She does not participate in, or contribute to, the definition of the Guaranteed Time Programme of the CHEOPS mission through which observations described in this paper have been taken, nor to any aspect of target selection for the programme. 
K.W.F.L. was supported by Deutsche Forschungsgemeinschaft grants RA714/14-1 within the DFG Schwerpunkt SPP 1992, Exploring the Diversity of Extrasolar Planets. 
This work was granted access to the HPC resources of MesoPSL financed by the Region Ile de France and the project Equip@Meso (reference ANR-10-EQPX-29-01) of the programme Investissements d'Avenir supervised by the Agence Nationale pour la Recherche. 
PM acknowledges support from STFC research grant number ST/M001040/1. 
GyMSz acknowledges the support of the Hungarian National Research, Development and Innovation Office (NKFIH) grant K-125015, a a PRODEX Experiment Agreement No. 4000137122, the Lend\"ulet LP2018-7/2021 grant of the Hungarian Academy of Science and the support of the city of Szombathely. 
V.V.G. is an F.R.S-FNRS Research Associate. 
NAW acknowledges UKSA grant ST/R004838/1. 
This paper includes data collected by the TESS mission. Funding for the TESS mission is provided by the NASA's Science Mission Directorate.
\end{acknowledgements}


\bibliographystyle{aa}
\bibliography{references}

\begin{appendix}
\section{\cheops\ observations of \keltb\ }

\begin{figure*}
    \centering
    \includegraphics[width=.49\hsize]{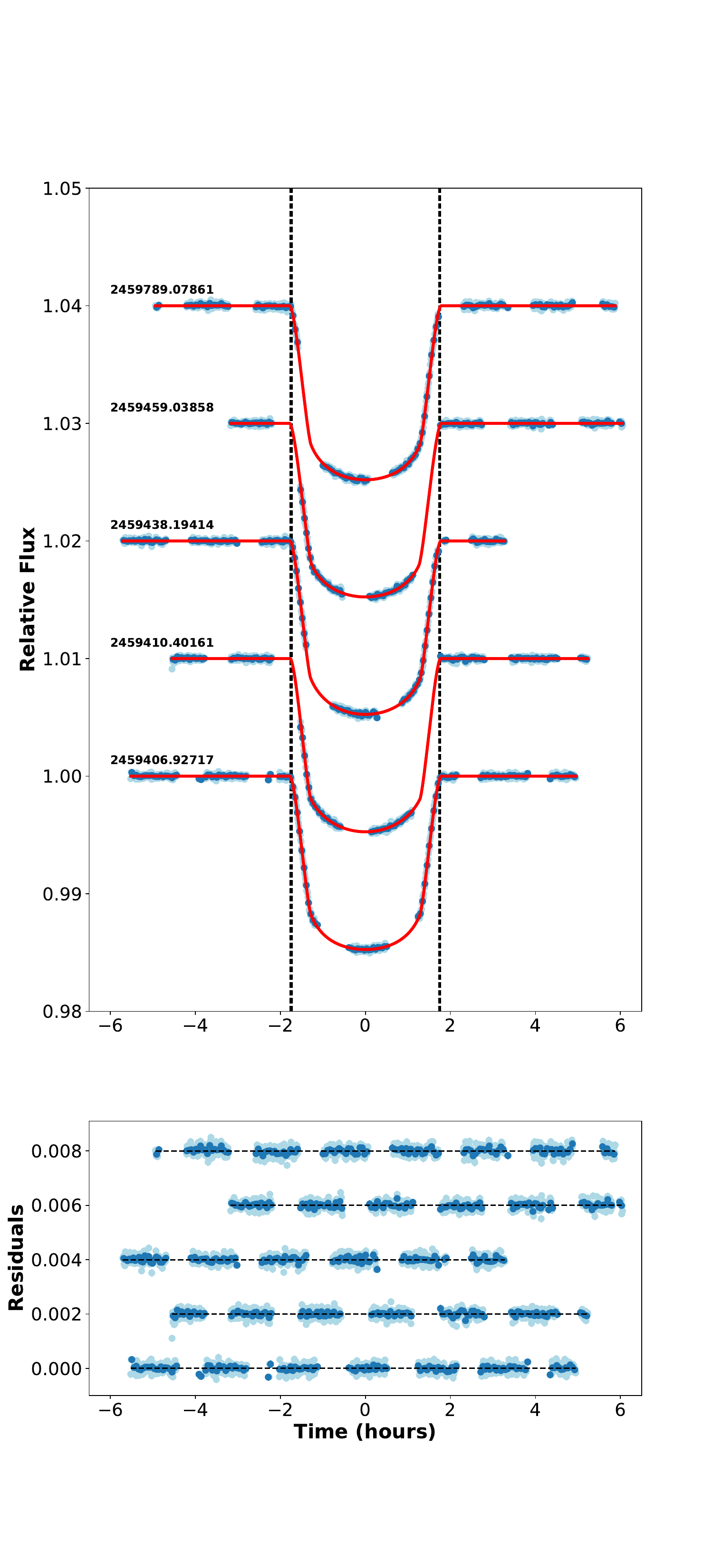}
    \includegraphics[width=.49\hsize]{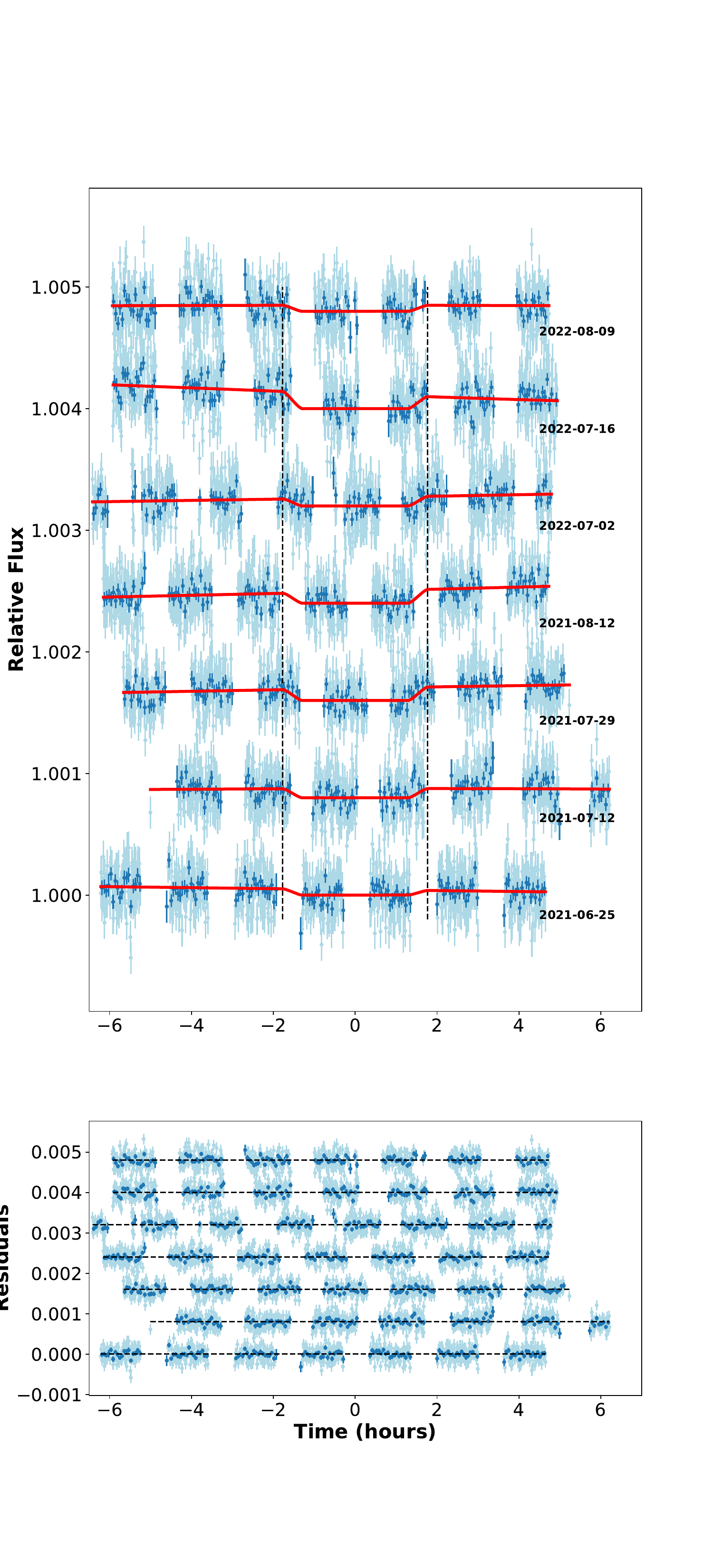}
    \caption{Individual \cheops\ transits (\textit{left}) and occultations (\textit{right}) of \keltb. The solid red line is the best-fit model, while the blue and light blue points represent the binned and unbinned light curves, respectively. The numbers in the figure represent the BJD time stamp of individual transit epochs. The residuals are shown in the bottom panel.} \label{fig:bestfits_cheops}
\end{figure*}

\begin{figure*}
\centering
\includegraphics[width=0.9\linewidth]{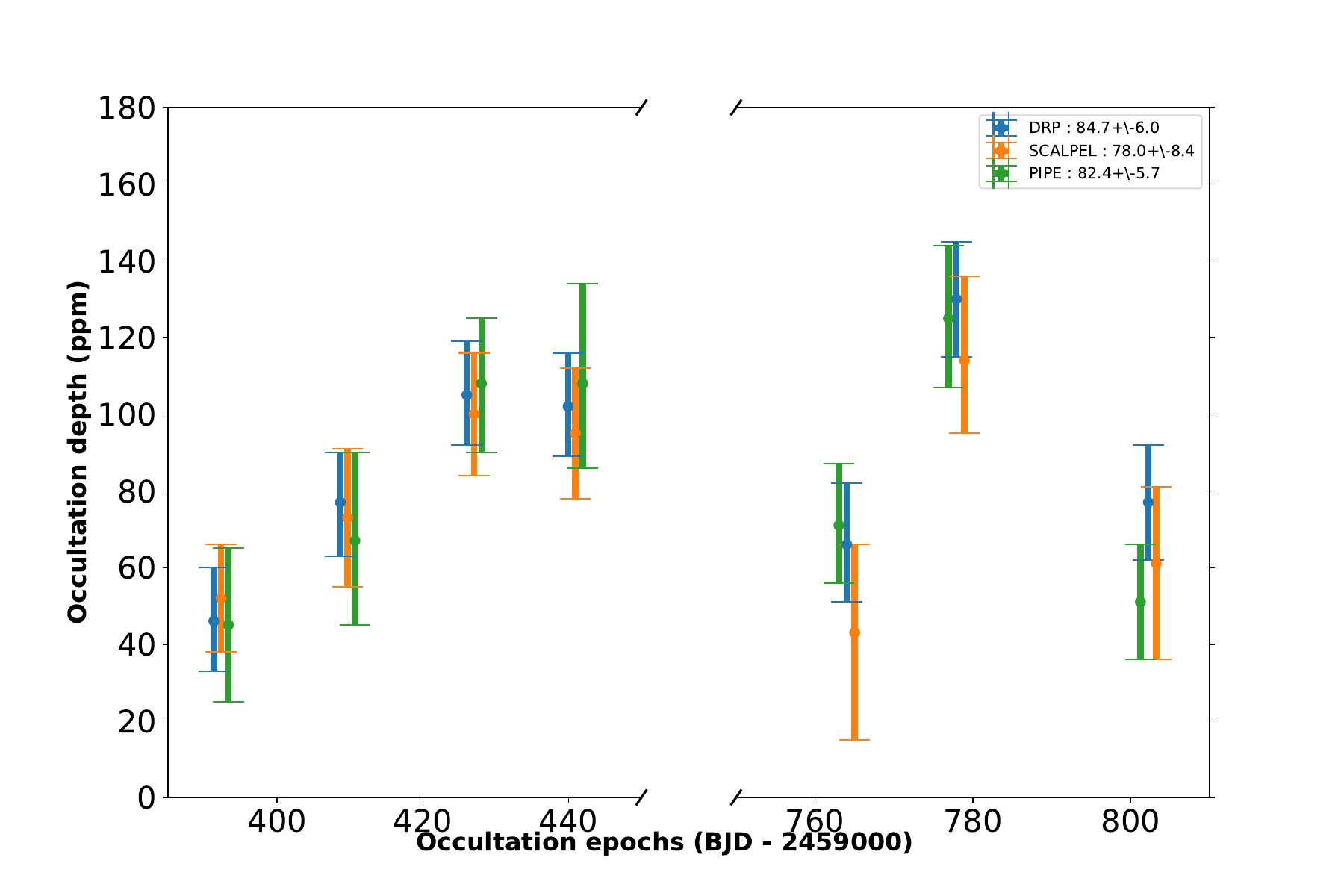}
\caption{\label{app:depths}Occultation depth measurements of \keltb\ of the seven \cheops\ observations. The three sets of measurements correspond to the three data extraction techniques discussed in Sect.~\ref{sec:observations}}
\label{fig: ComparativeVariability}
\end{figure*}

\begin{table*}
\caption{Best-fit parameters of the individual transit fits of \cheops}             
\label{tab: Transits_cheops}      
\centering          
\begin{tabular}{l c c c c c c r}    
\hline\hline       

Visit ID &       $T_{\mathrm{epoch}}$     &    $R_{p}/R_{\star}$    & $b$ & $\rho_{\star}$ [solar] & $q_1$ & $q_2$ &\\
\hline
\cr
TG000401 & 2459406.927174$\pm$0.000024 & 0.11573$^{+0.00030}_{-0.00031}$ & 0.516$\pm$0.007  &  0.458$\pm$0.005 & 0.278$^{+0.053}_{-0.047}$ & 0.289$^{+0.113}_{-0.094}$ & \\
\cr
TG000402 & 2459410.4016$^{+0.0366}_{-0.0393}$ & 0.11571$^{+0.00033}_{-0.00032}$ & 0.504$\pm$0.016  &  0.472$\pm$0.016 & 0.231$^{+0.028}_{-0.026}$ & 0.36$^{+0.063}_{-0.057}$ & \\
\cr
TG0001001 & 2459438.194141$^{+0.000032}_{-0.000033}$ & 0.11509$^{+0.00039}_{-0.00040}$ & 0.501$\pm$0.011  &  0.471$\pm$0.007 & 0.301$^{+0.047}_{-0.042}$ & 0.268$^{+0.067}_{-0.058}$ & \\
\cr
TG0001002 & 2459459.038579$^{+0.000251}_{-0.000244}$ & 0.11588$^{+0.00061}_{-0.00060}$ & 0.527$^{+0.027}_{-0.031}$  &  0.452$^{+0.031}_{-0.028}$ & 0.259$^{+0.041}_{-0.038}$ & 0.300$^{+0.082}_{-0.067}$  & \\
\cr
TG000801 & 2459789.078614$\pm$0.00044 & 0.11587$^{+0.00032}_{-0.00037}$ & 0.502$\pm$0.003  &  0.467$^{+0.029}_{-0.025}$ & 0.310$^{+0.037}_{-0.035}$ & 0.255$^{+0.052}_{-0.057}$  & \\
\cr

\hline                  
\end{tabular}\label{tab: transits_individual}
\end{table*}

\newpage
\section{\tess\ observations of \keltb\ }

\begin{table*}
\centering
\caption{{\tess}: Best-fit transit and phase curve free parameters of {\keltb}. }
\label{tab: TESS_sectorwise_phase_curve}
\begin{tabular*}{\linewidth} {@{\extracolsep{\fill}} lcccc}
\hline
Parameter   & Sector 14     & Sector 40    & Sector 41    & Sector 54  \\

\hline
\cr
$T_{\mathrm{epoch}}$ [TJD]  & $1694.736671\pm0.000062$ & $2403.453023\pm0.000038$     & $2434.720013\pm42$ & $2771.707828\pm0.000083$ \\
\cr
Period* [d]         & $3.474078^{+0.000029}_{-0.000030}$      & $3.474106\pm0.000016$ & $3.474086\pm0.000018$ &  $3.474101^{+0.000018}_{-0.000019}$  \\
\cr
$\rho_{\star}$ [solar]       & $0.470\pm0.006$      & $0.458\pm0.006$ & $0.461\pm0.007$ & $0.454\pm0.007$ \\
\cr
$R_{p}/R_{\star}$ [$\%$]         & $11.581^{+0.036}_{-0.039}$  & $11.569^{+0.026}_{-0.027}$ & $11.549\pm0.030$ & $11.603^{+0.026}_{-0.031}$ \\
\cr
$b$         & $0.508^{+0.008}_{-0.009}$   & $0.508\pm0.008$ & $0.513\pm0.009$ & $0.525^{+0.008}_{-0.009}$\\
\cr
$q_1$        & $0.146^{+0.042}_{-0.032}$      & $0.126^{+0.021}_{-0.019}$ & $0.146^{+0.024}_{-0.022}$ & $0.125^{+0.024}_{-0.019}$\\
\cr
$q_2$         & $0.268^{+0.099}_{-0.10}$     & $0.427^{+0.102}_{-0.086}$ & $0.326^{+0.0.091}_{-0.078}$  & $0.393^{+0.107}_{-0.098}$\\
\cr
$A_{\mathrm{day}}$  [ppm]       & $107\pm30$      & $102\pm22$ & $149\pm34$ & 141$^{+43}_{-46}$\\
\cr
$\log(h)$ & $-6.44^{+0.79}_{-0.35}$   & $-8.13^{+0.13}_{-0.11}$ & $-8.06^{+0.09}_{-0.08}$ & $-8.52^{+0.10}_{-0.09}$\\
\cr
$\log(l)$ & $2.57^{+0.60}_{-0.46}$     & $-0.58^{+0.13}_{-0.12}$   & $-1.34\pm0.09$  & $-1.56^{+0.16}_{-0.15}$ \\
\cr
Jitter [ppm]         &  $287\pm5$     & $178\pm3$ & $177\pm4$ & $171\pm5$ \\
\cr
\hline
\end{tabular*}
\end{table*}

\begin{figure*}
    \centering
    \includegraphics[width=.49\hsize]{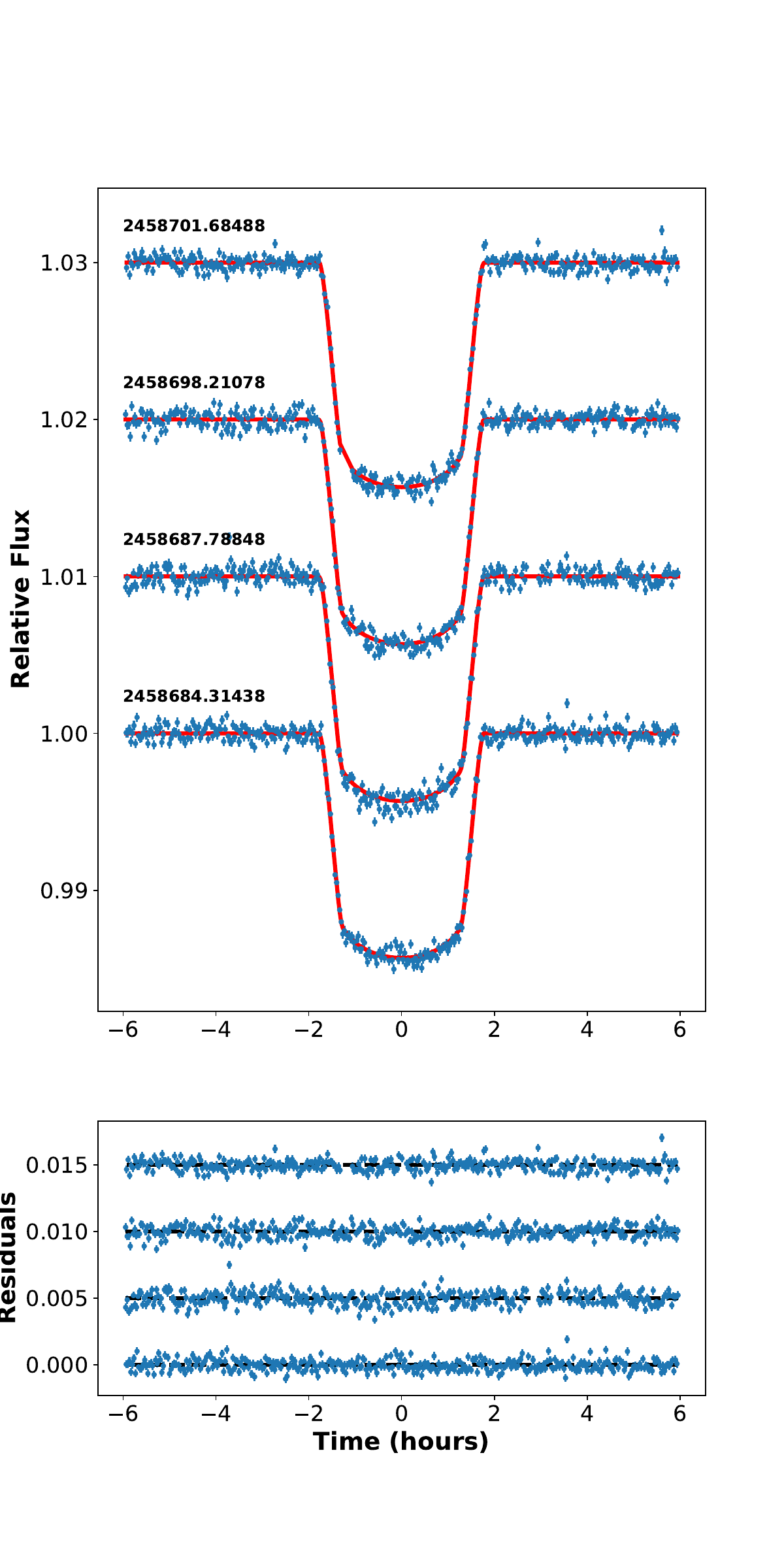}
    \includegraphics[width=.49\hsize]{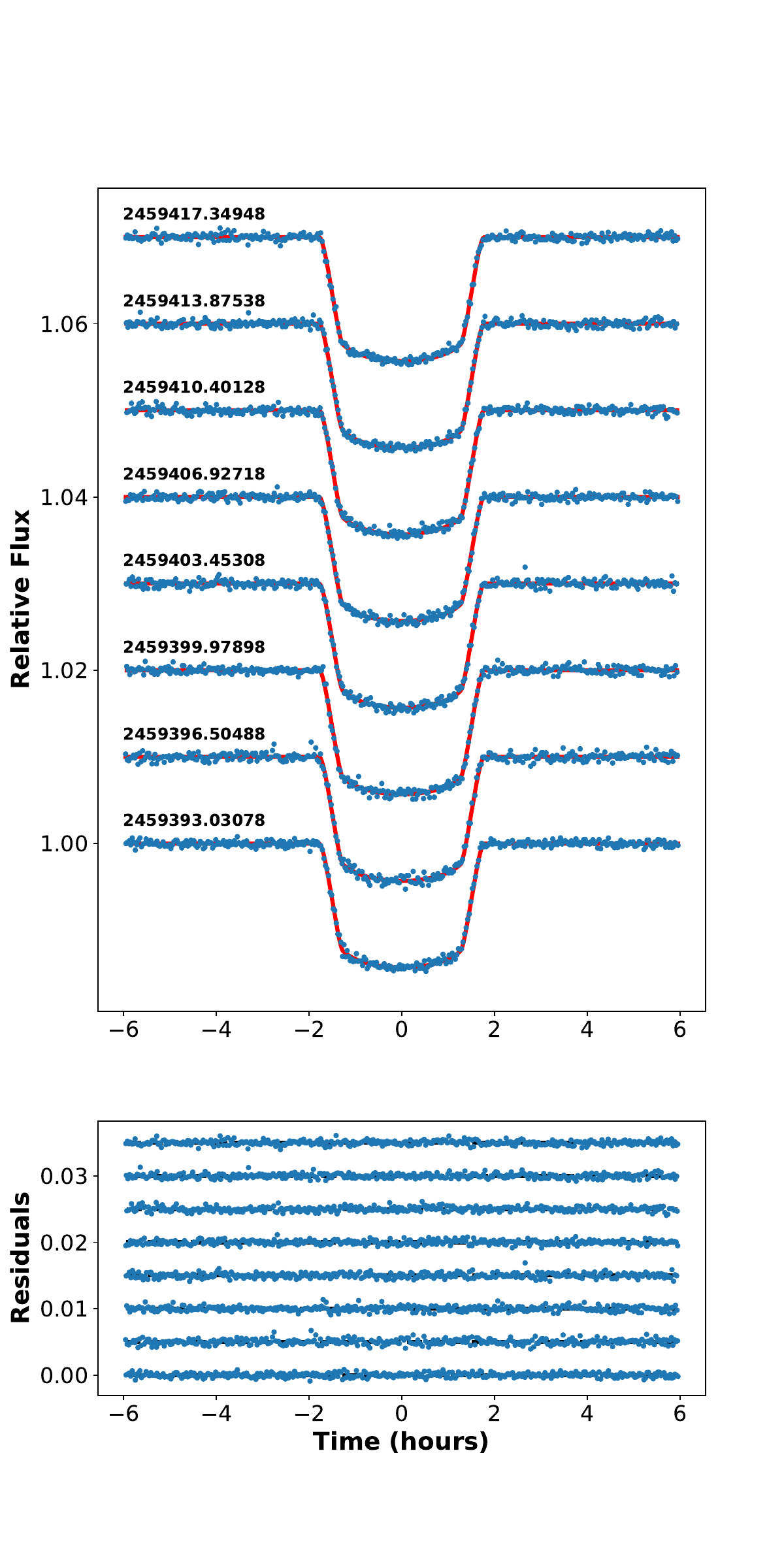}
    \caption{Individual transits of \keltb\ observed with \tess. The solid red line is the best-fit model, while the blue and light blue points represent the binned and unbinned light curves respectively. The numbers represent the BJD time stamp of individual transit epochs. The residuals are shown in the bottom panel. \textbf{\textit{Left: }}All transits of sectors 14. \textbf{\textit{Right: }}All transits of sector 40.} \label{fig:bestfit_transits_sector_14_40}
\end{figure*} 
    
\begin{figure*}
    \centering
    \includegraphics[width=.49\hsize]{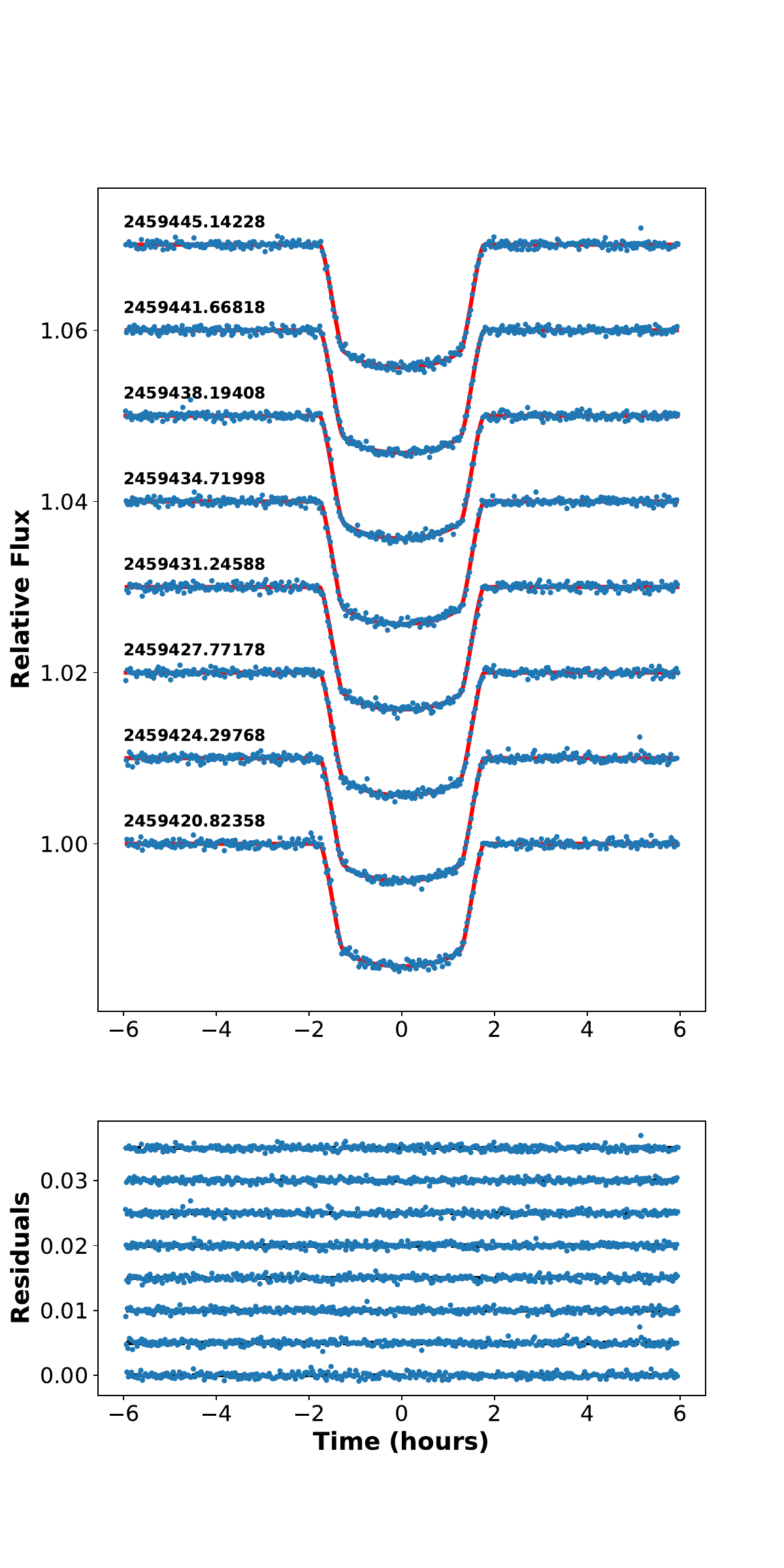}
    \includegraphics[width=.49\hsize]{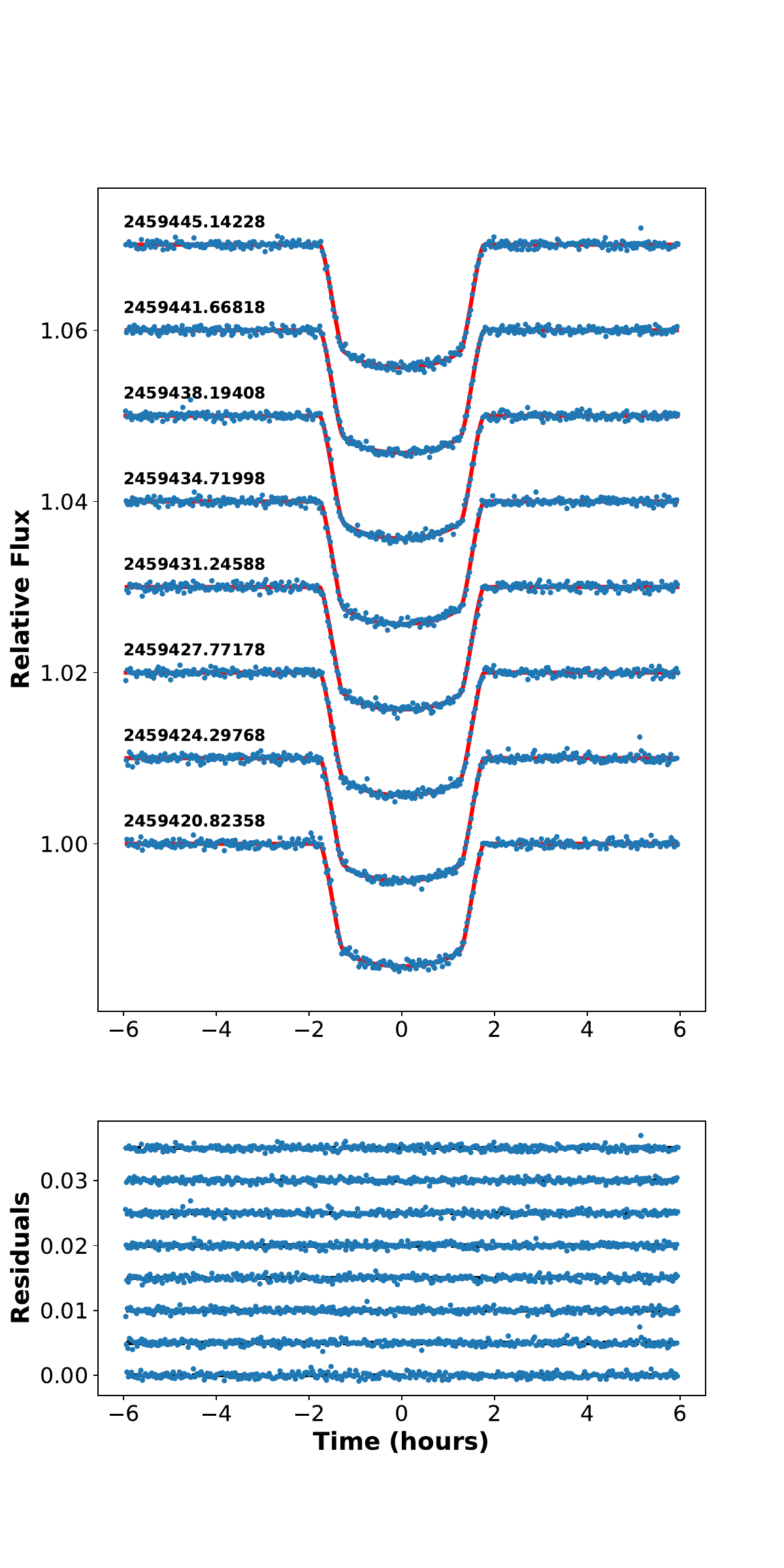}
    \caption{Same as in Fig. \ref{fig:bestfit_transits_sector_14_40}, but for sectors 41 and 54.}  \label{fig:bestfit_transits_sector_41_54}
\end{figure*}     

\begin{figure*}
\centering
\includegraphics[width=0.9\linewidth]{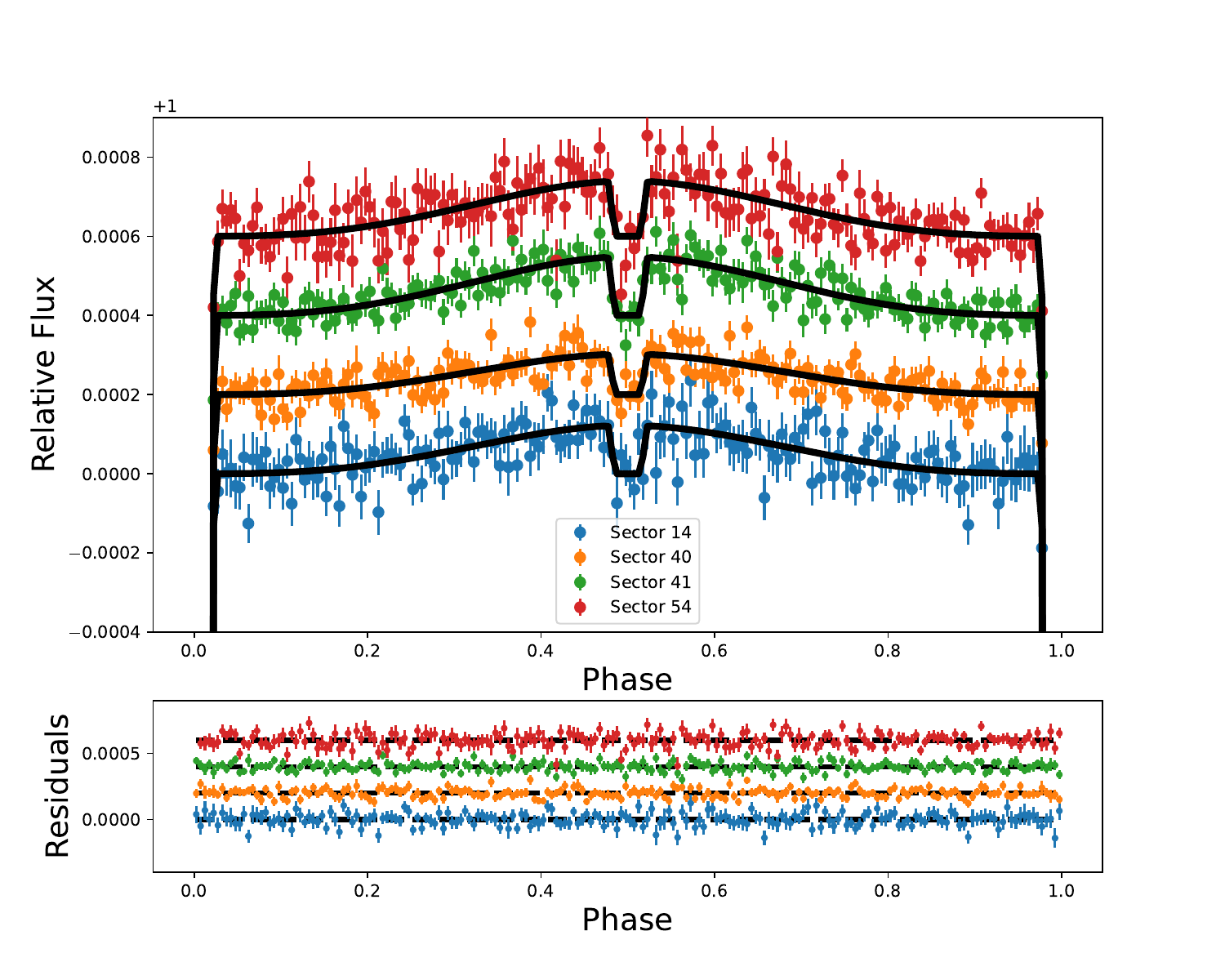}

\caption{Individual sector-wise \tess\ phase curves of \keltb. The variation in the occultation depth is somewhat apparent between Sectors 40 and 41. The sector-wise occultation depths are $123^{33}_{34}$, 102$\pm$34, 149$\pm$35, and $141^{+43}_{-46}$ ppm for sectors 14, 40, 41, and 54 ppm, respectively. The corresponding residuals are in the bottom panel.} 
\label{fig: TESS_sector-wise}
\end{figure*}

\begin{figure*}
    \centering
    \includegraphics[width=0.95\textwidth]{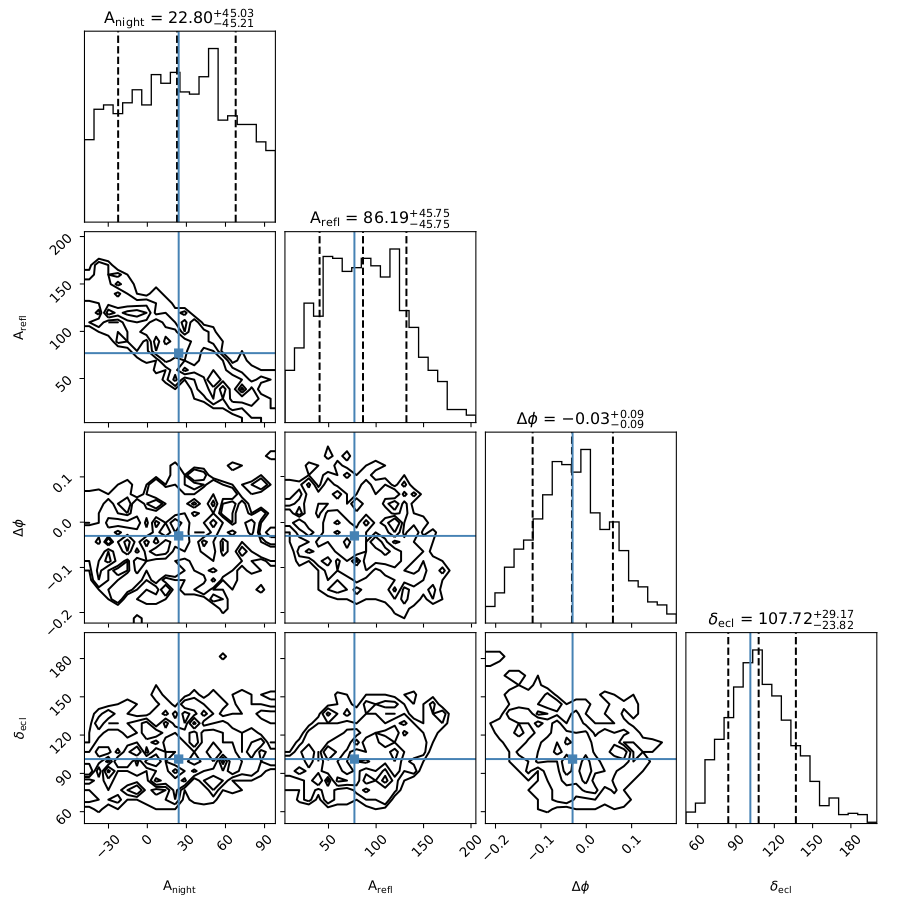}
    \caption{ Posteriors of the model that includes the phase offset and the nightside flux as free parameters for \tess\ Sector 40.  $\Delta\phi$ varies between 0 and 1 ($360^{\circ}$). Both $A_{night}$ and $\Delta_\phi$ are consistent with zero, and the occultation depth measurement is the same as that measured when they are fixed to zero (see Table~\ref{tab: TESS_sectorwise_phase_curve}). 
    } \label{fig: A_night_Ph_offset}
\end{figure*}

    \section{Derived parameters}

    \begin{figure*}[t]
    \centering
    \includegraphics[width=0.8\linewidth]{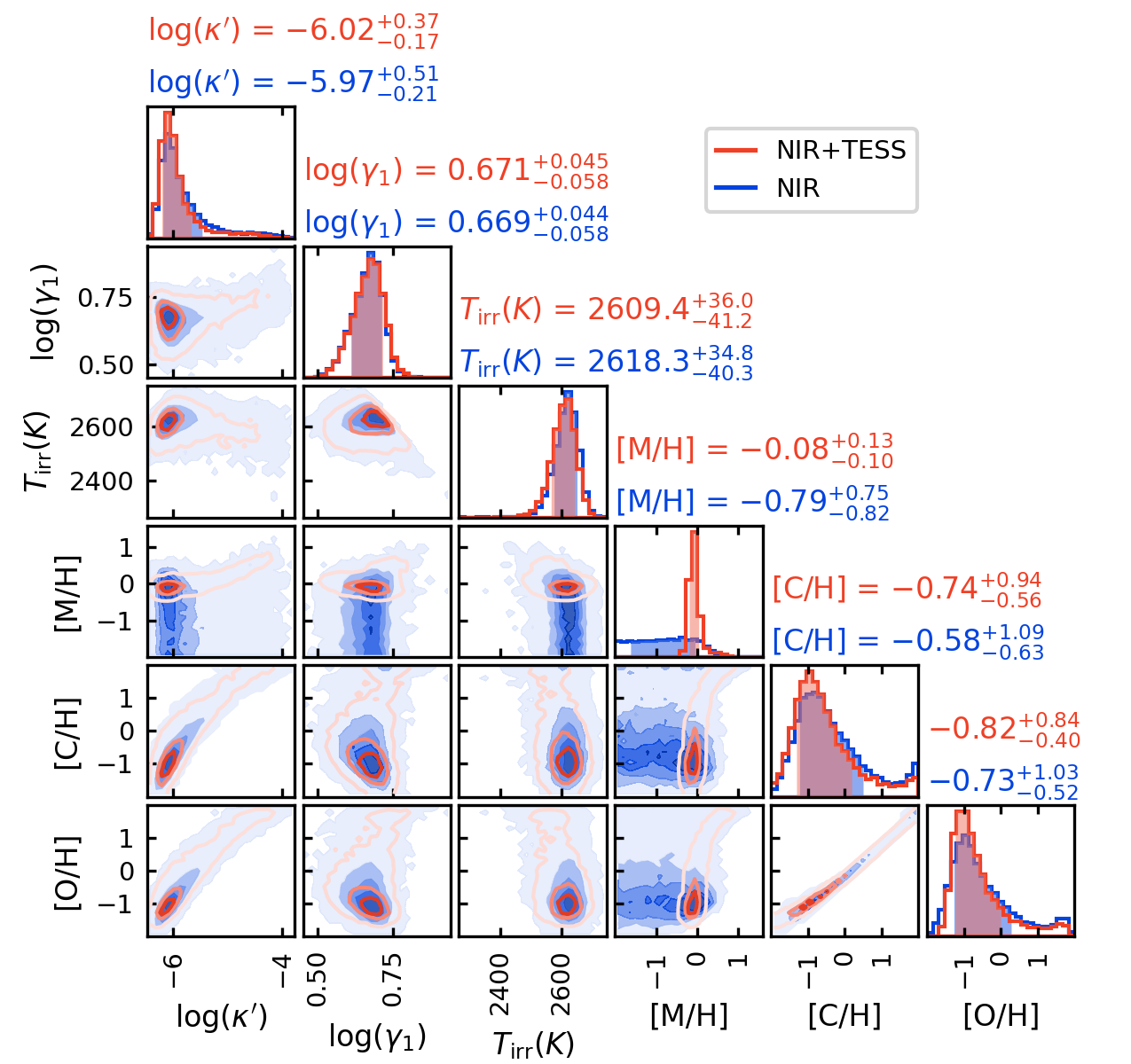}
    \caption{Posterior distributions for our {\keltb} atmospheric
      retrievals constrained by the NIR occultations ({\HST} and
      {\Spitzer}, blue) and the NIR+TESS occultations (red). The reported
      values at the  top of each marginal posterior denote the median of the
      distribution and the boundaries of the central 68\% percentile
      (i.e. the 1$\sigma$ credible intervals).  The shaded areas in the
      marginal histograms denote the span of the credible intervals.}
    \label{fig:retrieval_posteriors}
    \end{figure*}

    \begin{figure*}[h!]
    \centering
    \includegraphics[width=0.95\linewidth]{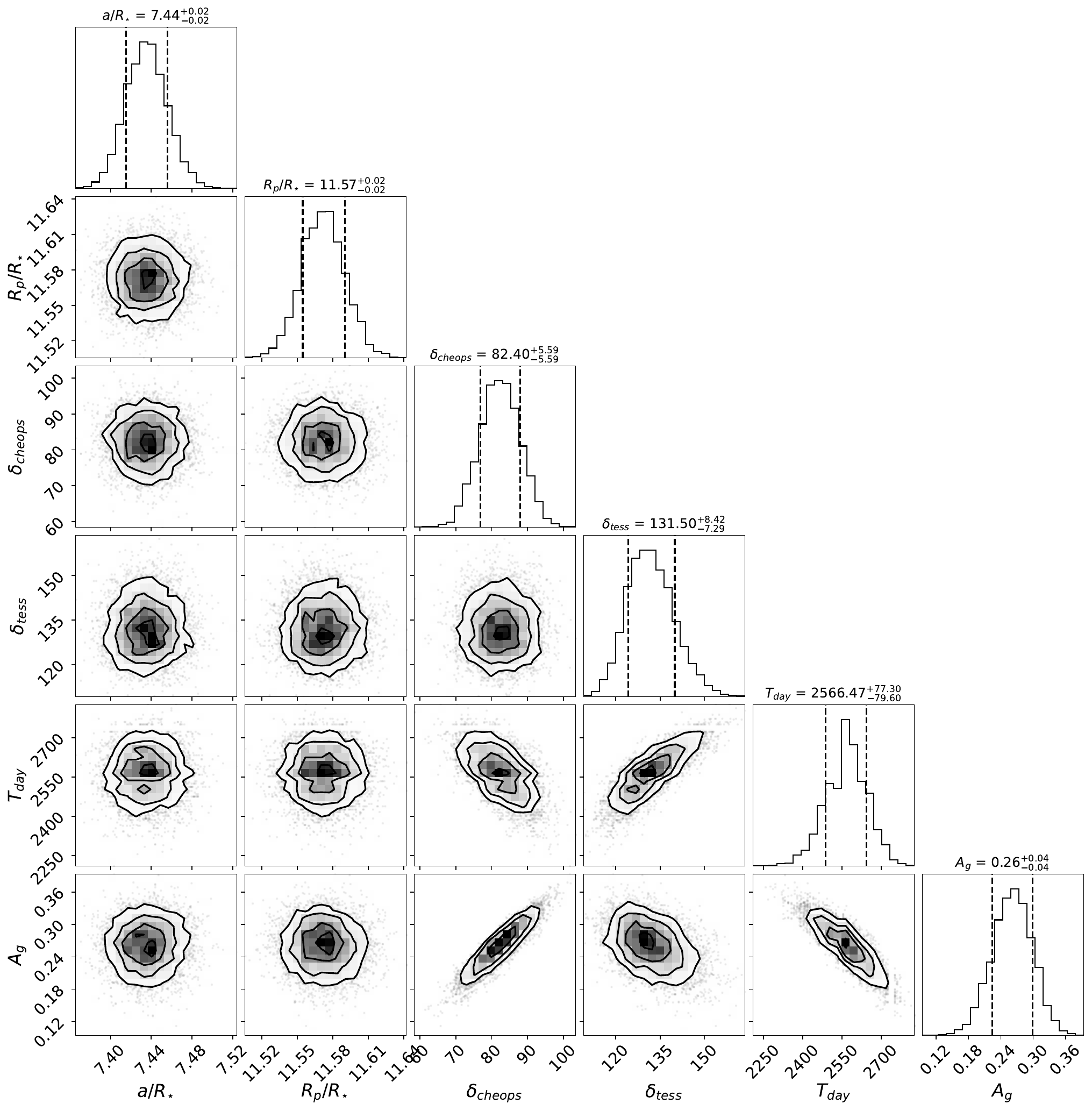}
    \hspace{1.0cm}
    \caption{Posterior distribution of the \cheops\ and \tess\ eclipse depths in ppm along with the  planetary fit parameters $R_{p}/R_{\star}$ [$\%$] and a/$R_{\star}$ from the light curve fitting are used to create the posteriors on the dayside brightness temperature and geometric albedo (for reference, see Sect.~\ref{sec: discussion}).} 
    \label{fig: Ag_Tday_posterior}
    \end{figure*}
\end{appendix}

\end{document}